\documentclass[11pt,a4paper]{article}
\usepackage{amsmath,amssymb}
\usepackage{epsfig,graphicx}
\usepackage{subfigure}
\usepackage{rotating}
\usepackage{bm}
\usepackage{color}

\newlength{\wth}
\newcommand{\startappendix}{
\setcounter{section}{0}
\renewcommand{\thesection}{\Alph{section}}}

\def\CH{{\cal H}}
\def\CS{{\cal S}}

\def\sst{\scriptscriptstyle}

\def\Dbarslash{\,\,{\raise.15ex\hbox{/}\mkern-12mu {\bar\D}}}
\def\Dslash{\,\,{\raise.15ex\hbox{/}\mkern-12mu \D}}
\def\delslash{\,\,{\raise.15ex\hbox{/}\mkern-9mu \partial}}
\def\delbarslash{\,\,{\raise.15ex\hbox{/}\mkern-9mu {\bar\partial}}}

\def\D{{\cal D}}
\def\Dbarslash{\,\,{\raise.15ex\hbox{/}\mkern-12mu {\bar\D}}}
\def\delslash{\,\,{\raise.15ex\hbox{/}\mkern-9mu \partial}}
\def\Dslash{\,\,{\raise.15ex\hbox{/}\mkern-12mu \D}}

\def\={\, =\, }
\def\+{\, +\, }
\def\-{\, -\, }

\newcommand{\be}{\begin{equation}}
\newcommand{\ee}{\end{equation}}
\def\bea{\begin{eqnarray}}
\def\eea{\end{eqnarray}}

\definecolor{saabeer}{rgb}{0,1,0}

\definecolor{durbeer}{rgb}{1,0,0}

\definecolor{durbeer2}{rgb}{0.8,0,0.5}

\begin{document}
\date{\mbox{ }}
\title{{\normalsize  IPPP/09/81; DCPT/09/162;  DAMTP-2009-64\hfill\mbox{}\hfill\mbox{}}\\
\vspace{2.5cm} \LARGE{\textbf{
Phenomenology of Pure General Gauge Mediation}}}
\author{Steven Abel$^a$, Matthew J. Dolan$^b$, Joerg Jaeckel$^a$ and  Valentin V. Khoze$^a$\\[2ex]
\small{$^a$\em Institute for Particle Physics Phenomenology,}\\
\small{\em Durham University, Durham DH1 3LE, United Kingdom}\\[2ex]
\small{$^b$ \em DAMTP, CMS, University of Cambridge, Wilberforce Road,}\\
\small{\em Cambridge, CB3 0HE, United Kingdom}}  
\date{}
\maketitle

\vspace{3ex}

\begin{abstract}
\noindent
We investigate the phenomenology of general gauge mediation in the MSSM. We apply
the strict definition of gauge mediated SUSY-breaking where $B_\mu$ is
generated only through gauge interactions, and as a result is very close to
zero at the messenger scale. In this setup $\tan\beta$ is a prediction rather
than an input. The input parameters are independent scales for
the gaugino masses, the scalar masses and the messenger mass
in accord with general gauge mediation.
We investigate the spectra, the constraints on the parameter space from direct searches
and indirect observables, as well as fine-tuning.
The favoured region of parameter space includes and interpolates between non-split
and mildly split SUSY, characteristic of ordinary gauge mediation
and direct gauge mediation models, respectively.

\end{abstract}

\newpage

\tableofcontents

\section{Introduction}
One of the appealing properties of gauge mediated supersymmetry (SUSY) breaking
is its restricted parameter space. This contrasts with
gravity mediation where the parameter space
is of high dimensionality.
However, it has become evident that the parameter space of gauge mediation is
larger than we used to imagine.
In particular, the authors of~\cite{Meade:2008wd} have introduced
a novel framework suitable for discussing and analysing
very general models of gauge mediation in a model-independent way.
It is now timely to undertake a phenomenological study
of general gauge mediation to complement those of gravity mediation and this is
the primary motivation of this paper (as well as recent work in Refs.~\cite{Carpenter:2008he,Rajaraman:2009ga}).

The General Gauge Mediation (GGM) paradigm~\cite{Meade:2008wd} is
defined by the requirement that the Minimal Supersymmetric Standard
Model (MSSM) becomes decoupled from the hidden SUSY-breaking sector
in the limit where the three MSSM gauge couplings $\alpha_{i=1,2,3}$
are set to zero. Since no other parameters participate
in the coupling of the two sectors, this strict interpretation of gauge mediation
can be called `general pure gauge mediation' or pure GGM. This framework is broad
enough to include everything from weakly coupled models with
explicit messengers to strongly coupled theories with direct
mediation.

The main free parameters in this setup are the gaugino and scalar masses as well as
the messenger scale. For simplicity we restrict ourselves in this work to a single effective
scale $\Lambda_G$ for the gaugino masses and a single scale $\Lambda_S$ for the
scalars\footnote{We do not split the scale
for the different gauge representations as was
done in~\cite{Carpenter:2008he,Rajaraman:2009ga}.}.
Thus at the messenger scale $M_{mess}$ the soft supersymmetry breaking gaugino masses are
\begin{equation}
\label{gauginosoft}
M_{\tilde{\lambda}_i}(M_{mess}) =\, k_i \,\frac{\alpha_i(M_{mess})}{4\pi}\,\Lambda_G
\end{equation}
where $k_i = (5/3,1,1)$, $k_i\alpha_i$ (no sum)
are equal at the GUT scale and $\alpha_i$ are the gauge coupling constants.
The scalar mass squareds are
\begin{equation}
\label{scalarsoft}
m_{\tilde{f}}^2 (M_{mess}) =\, 2 \sum_{i=1}^3 C_i k_i \,\frac{\alpha_i^2(M_{mess})}{(4\pi)^2}\, \Lambda_S^2
\end{equation}
where the $C_i$ are the quadratic Casimir operators of the gauge groups.
Ordinary gauge mediation scenarios (see Ref.~\cite{Giudice:1998bp} for a review) live on the restricted
parameter space $\Lambda_G\simeq\Lambda_S$.

Outside the confines of \emph{ordinary} gauge mediation the
parameter space is populated by many models
that
predict different values of the ratio of gaugino to scalar masses, $\Lambda_G/\Lambda_S$. In
models with explicit messengers one expects
this ratio to be close
to one, while for direct mediation models the gaugino masses are
often suppressed relative to the scalar
masses~\cite{Izawa:1997gs,Kitano:2006xg,Csaki:2006wi,Abel:2007jx,Abel:2007nr,Abel:2008gv}.
Recently, hybrid models have been constructed which interpolate
between these two cases~\cite{Abel:2009ze}. It is also possible to
achieve values $\Lambda_G/\Lambda_S>1$ by increasing the ``effective number of
messengers'' in the context of extraordinary gauge mediation
models~\cite{Cheung:2007es}.
Indeed we argue that the set of models defined by $\Lambda_G$, $\Lambda_S$ and $M_{mess}$
are the gauge mediation equivalent to the canonical mSUGRA (or Constrained MSSM)
scenario, with $\Lambda_G$ and $\Lambda_S$ playing the role of the parameters $m_{1/2}$ and
$m_0$ in those models.

 With such a plethora of possibilities suddenly available, it is therefore
important to determine if any region
in this parameter space is favoured by experimental data.
Accordingly, in this paper we will confront the full $\Lambda_G$, $\Lambda_S$ and
$M_{mess}$ parameter space with a number of measured
observables
in order to provide direction for model building and investigate
expected LHC signals.

Before we proceed to the phenomenology we outline our approach to the
supersymmetry breaking in the Higgs sector.
Pure General Gauge Mediation on its own does not generate
the $\mu$-parameter appearing in the effective Lagrangian,
\be
\label{mudef}
{\cal L}_{eff}\supset\int d^2 \theta \,\, \mu \,\CH_u\CH_d \ ,
\ee
where the Higgs superfields are denoted by $\CH$ and
their scalar components are $H$.
The phenomenologically required value of $\mu$ is roughly of the order of
the electroweak scale and as usual will be determined in our
analysis from the requirements of electroweak symmetry breaking.

In addition to the supersymmetric interaction \eqref{mudef},
the Higgs-sector effective Lagrangian also includes
soft supersymmetry-breaking terms. All of the latter must be
generated by the SUSY-breaking sector,
since there would be little merit in
a model of dynamical SUSY-breaking which generates only a
subset of the SUSY-breaking terms in the effective SM Lagrangian.
There are quadratic terms
\be
\label{quaddef}
m_u^2 |H_u|^2 + m_d^2 |H_d|^2 +(B_\mu H_uH_d + c.c.)~,
\ee
as well as cubic $a$-terms
\be
\label{Atermsdef}
a_u^{ij} H_u Q^i \bar u^j + a_d^{ij} H_d Q^i \bar d^j + a_L^{ij} H_d L^i \bar E^j~,
\ee
in the MSSM. As is well-known, a phenomenologically acceptable electroweak
symmetry breaking in the supersymmetric SM occurs
if $\mu^2$ and the soft masses in \eqref{quaddef} at the low scale
(i.e. the electroweak scale)
are of the same order, $\mu^2 \sim B_\mu \sim m_{soft}^2\sim M^2_{W}$.

In a strict interpretation of gauge mediation, where we have no direct couplings
of the SUSY-breaking sector to the Higgs sector, we have $B_{\mu}\approx 0$ at the
messenger scale.
In Section~\ref{sec:bmu} we will argue that more generally it is
indeed natural to have negligibly
small input values for $B_\mu$ at the messenger scale, $B_\mu \ll \mu^2 \sim m_{soft}^2$.
From this starting point, i.e. taking $B_{\mu}\approx 0$ at the
high scale $M_{mess}$, a quite small but perfectly viable value of $B_\mu$ is then generated
radiatively at the electroweak scale
\cite{Rattazzi:1996fb,Babu:1996jf}.

We then use the measured value of the mass of the Z-boson to predict
values of $\tan\beta$ and $\mu$ from the requirement of electroweak
symmetry breaking. Since it is $B_\mu$ which is responsible for
communicating the vev of $H_u$ to $H_d$, this implies that the ratio
of these two vevs, $\tan\beta$, will be large (between $15$ and
$65$). This is in contrast to the common approach where $\tan\beta$
is taken as an arbitrary input and $B_{\mu}$ at the high scale is
obtained from it. For us $B_{\mu}$ (rather than $\tan\beta$) is the
fundamental quantity.

The fact that $\tan\beta$ is expected to be large
results in distinctive phenomenological features.
For example, it is well known that certain Yukawa couplings in the MSSM
are enhanced at large $\tan\beta$, leading to significant
supersymmetric contributions to rare flavour-changing branching ratios.
SUSY loop effects can also explain the
discrepancy between the measured value of the anomalous magnetic
moment of the muon and its predicted value in the Standard Model.
Accordingly, precision measurements give us an
opportunity indirectly to constrain the GGM parameter space.

In the following Section we discuss the logical possibilities
for $\mu$ and $B_{\mu}$ with reference to their generation
in SUSY-breaking GGM models.
The detailed phenomenological study is
given in Section~\ref{sec:pheno} with the spectra of the best-fit
points in Appendix~\ref{sec:Spectra}.
A complementary analysis of the parameter space in
terms of the fine-tuning necessary to break electroweak symmetry
is presented in Section~\ref{sec:finetuning}.

\section{General Gauge Mediation and $\mu \, / \, B_{\mu}$}\label{sec:bmu}

In the context of gauge mediation, a typical coupling of the Higgs fields to the
SUSY-breaking sector leads to
$B_\mu \gg \mu^2$ which is phenomenologically unacceptable. This is often referred to as the $\mu/B_\mu$-problem of gauge mediation
which can be reconsidered in the context of the extended GGM
construction of \cite{Komargodski:2008ax} with SUSY-breaking sectors involving
more than a single scale.
Earlier approaches using particular gauge mediation models were constructed previously in
Refs.~\cite{Dvali:1996cu,Dine:1996xk,Yanagida:1997yf,Dimopoulos:1997je,Langacker:1999hs,Hall:2002up,Giudice:2007ca,Liu:2008pa,Csaki:2008sr}.

In this section we will describe the conceptual reasons that
compel us to take $B_{\mu}\approx 0$ at the messenger scale.
A fuller discussion of the theoretical underpinnings as well as some specific
model building examples are given in Appendix \ref{app:bmu}.
We will also review the RG behaviour which generates a small but
phenomenologically viable value of $B_{\mu}$ at the weak scale.

Very generally we can think of three logical possibilities for $\mu$:

\noindent(a) Being part of the superpotential, $\mu$ a priori has
nothing to do with SUSY-breaking and in particular with the
gauge-mediation mechanism. In this scenario $\mu$
in \eqref{mudef} appears as a tree-level parameter  from the GGM perspective. The reason why
$\mu \ll M_{Pl}$ would then have to be addressed in a way
decoupled from any SUSY-breaking mechanism. In other words, some
SUSY-preserving new physics would have to generate appropriate
$\mu$ dynamically by giving VEVs to appropriate fields; $\mu$
is not a problem of gauge mediation and $B_{\mu}$ is roughly zero
at the messenger scale. This is the pure GGM setting.

\noindent (b) The second possibility is that $\mu$ is generated
by the SUSY-breaking sector such that $\mu \propto M$,
where $M$ is a SUSY-breaking sector mass-scale,
which however is distinct from (and much higher than) the SUSY-breaking
scale, $M^2 \gg F$. In order to enable
such $\mu$-generation in the first place, one needs to extend the pure GGM of \cite{Meade:2008wd} to
include additional non-gauge
couplings between the Higgses and the fields of the SUSY-breaking sector.
This set-up corresponds to
the so-called two-scale models in the terminology of \cite{Komargodski:2008ax}.
The coupling to the Higgses then, as we discuss in Appendix~\ref{app:bmu}, automatically
induces a non-vanishing $B_{\mu}$.

\noindent (c) Finally, one could imagine generating $\mu$ by coupling Higgs fields
to a simple one-scale SUSY-breaking sector.
In this case $\mu \propto F/M$.
Here $F$ is the SUSY-breaking $F$-term and $M$ is the messenger mass.
$F/M$ is the single mass-scale characterising this scenario.
Again, $B_{\mu}$ is generated by the coupling of the
Higgses to the SUSY-breaking sector, and, as is well
known, it typically is unacceptably large and wrecks phenomenology (see Appendix~\ref{app:bmu}).

In the scenarios of case (a) the entire $\mu/B_\mu$-issue does not present a problem.
The $\mu$-parameter is generated by
high-energy physics distinct from SUSY-breaking and can be viewed
as a tree-level effect in the MSSM. At the same time $B_\mu$ is generated by normal gauge mediation at two loops
in the SM couplings. There are no additional effects on $B_\mu$ from the SUSY-preserving sector which has generated $\mu$.
In practice this amounts to taking $B_\mu \simeq 0$ at the (high) messenger scale
and treating $\mu$ alone as an a priori arbitrary input parameter which will be fixed (together with $\tan \beta$) to achieve
appropriate electroweak symmetry breaking at the low scale.
All models of the type (a) naturally fit within the class we have
phenomenologically analysed in Section~\ref{sec:pheno}.

The only other acceptable scenario -- case (b) can quite easily resolve
the $\mu/B_\mu$-problem, which is the
central result of \cite{Komargodski:2008ax}.
What happens is that the $\mu$-parameter is induced at a
high (SUSY-preserving) scale $M$ through non-gauge couplings
between the Higgses and the SUSY-breaking sector.
At the same time, these couplings also generate new contributions to the
SUSY-breaking $B_\mu$ and other soft parameters
(on top of the usual pure gauge-mediated effects). Since,
contrary to $\mu$, the soft terms must vanish when SUSY is restored,
they are generated at the SUSY-breaking scale $F\ll M^{2}$.

For the particular choice of parameters advocated in \cite{Komargodski:2008ax},
the resulting pattern (at the messenger scale) is
\be
B_\mu \sim \mu^2 \sim m^2_{soft}
\label{samesc}
\ee
As argued in Appendix~\ref{app:bmu} case (b)
can naturally generate a different pattern (at the messenger scale),
\be
B_\mu \ll \mu^2 \sim m^2_{soft}.
\label{diffsc}
\ee
In this case the models of type (b) are also included in the class we have
phenomenologically analysed in Section~\ref{sec:pheno}.

\subsection{The $B_{\mu}$ Parameter at Low Energy}
\label{sec:appB}
In order to perform the phenomenological analysis we need to compute the
weak scale value of $B_{\mu}$ in Eq.~\eqref{diffsc} after its running down from the
messenger scale. In practice we will do this
numerically as for all the other parameters. However to get a feeling for the important
effects we will briefly review the analytic approximation, as
presented in Ref.~\cite{Giudice:1998bp}. It is convenient to work with the parameter
\begin{equation}
B=\frac{B_{\mu}}{\mu}
\end{equation}
which is of mass dimension one.
If at scale $M_{mess}$ the $B$ parameter is set
to an initial value $B_0$, then at a low scale \mbox{$Q (t= \ln(M_{mess}^2/Q^2))$}
it runs to be~\cite{Giudice:1998bp}
\begin{equation}
\label{bequation}
B(t) = B_0 + \left(H_4 - \frac{K_t}{2} H_2 \right) \Lambda_G + \delta B^{NLO}(t).
\end{equation}
In our approach we will, as we have already mentioned, take the initial value $B_{0}=0$.

The remaining terms in \eqref{bequation} which represent the running include
\[
H_4 = \sum_{r=1}^3 \frac{a_r^{\mu}}{b_r} k_r \frac{\alpha_r(0)- \alpha_r(t)}{4\pi}
\]
with $a_r^{\mu} = 2(C_r^{H_u}+ C_r^{H_d}) = (1,3,0)$, and $b_r = (11,1,-3)$ being
the beta function coefficients. Also,
defining
\begin{equation}
E = \prod_{r=1}^3 \left[ \frac{\alpha_r(0)}{\alpha_r(t)}\right]^{\frac{a_r}{b_r}}, \qquad
F = \int^t_{0} dt E
\end{equation}
where $a_r = w ( C_r^{Q_L} + C_r^{\tilde{t}_R} + C_r^{H_u}) = (13/9,3,16,3)$ one has
\begin{equation}
H_{2}=\frac{\alpha_X t_X}{4\pi} \left[ \frac{E}{F} \left( \frac{t}{t_X} -1 \right) + \frac{1}{F} - \frac{1}{t_X} + \sum_{r=1}^3 a_r \frac{\alpha(0)}{4\pi} \right]
\end{equation}
where $t_X = \ln (M^2_{mess} / M_X^2)$  and $\alpha_X$ are the unification scale and unification gauge coupling constant in a theory without messengers, and
\[
K_t = \frac{6F}{E} \frac{h_t^2(t)}{\left( 4\pi \right)^2}
\]

The next to leading order corrections are given by
\begin{equation}
\delta B^{(NLO)}(t) = - \frac{\alpha_s^2(t) h_t^2(t)}{8\pi^4} t \Lambda_G + \sum_{r=1}^3 a_r^{\mu} k_r
\frac{\alpha_r^2(t)}{\left( 4\pi \right)^2} \Lambda_G
\end{equation}
which is suppressed by an additional loop factor or a factor of $1/t$ with respect
to the leading order contribution.

The middle term in Eq.~\eqref{bequation} is the pure running effect and vanishes at the
messenger scale, so that
\begin{equation}
B(t=0)=B_{0}+\delta B^{(NLO)}(t=0).
\end{equation}
As we already noted the first term on the right hand side
is zero in our setup and the second term is precisely
the two-loop contribution to $B$ generated at the messenger scale.

The above gives the reader an analytical understanding of the
generation of $B$. As we have said the {\texttt SoftSUSY} program
uses the numerical integration of the RG equations, and this
includes additional contributions from down-type quarks, leptons and
all three families.

\section{Phenomenology}\label{sec:pheno}
Recent work on the phenomenology of GGM
includes Refs.~\cite{Carpenter:2008he,Rajaraman:2009ga} where $\tan\beta$ was taken
as a free input parameter and $B_\mu$ computed from it.
As explained in the Introduction, our approach is to take the fundamental parameter $B_\mu$
as an an input; we take the theoretically motivated value $B_\mu=0$ at the messenger scale.
A non-zero value of $B_{\mu}$ is
then generated radiatively (see below and also Section \ref{sec:appB}).
This predicts specific values for $\tan\beta$.

In the following we present constraints on the parameter space and details of the spectrum,
commenting on the implications for
SUSY searches at the LHC. We continue by discussing the implications of low energy
precision observables for our scenario and the preferred values of $\Lambda_G/\Lambda_S$.
All the results we discuss are for $\mu>0$.
(We have also done scans for $\mu<0$ and found that it is
phenomenologically disfavoured by the low-energy observables. We do not present results
for this case.)

To compute the spectrum from the
soft terms given in Eqs.~\eqref{gauginosoft} and \eqref{scalarsoft}
we use a version of \texttt{SoftSUSY 3.0.9}~\cite{Allanach:2001kg}
which we modified to accept $B_{\mu}$, rather
than $\tan\beta$, as an input. The value of $\tan\beta$ is set according to the
requirement that electroweak symmetry be correctly broken. We have
scanned over the range $ 3.0\times 10^4 \leq \Lambda_G \leq
3.0\times 10^6$~GeV in the gaugino masses and $1.0\times 10^3  \leq
\Lambda_S \leq 3.0 \times 10^6$~GeV in the scalar masses, where
the parameters $\Lambda_G$ and $\Lambda_S$ were
defined in Eqs.~\eqref{gauginosoft} and \eqref{scalarsoft} (for $M_{mess}=10^6$~GeV we alter
the upper bounds on $\Lambda_{G,S}$ to $1.0\times 10^6$~GeV). We have chosen a
smaller lower bound in the scalar mass parameter $\Lambda_S$ as this
is not ruled out from a purely phenomenological point of view due to
significant effects of the gaugino masses in the running of the
scalar masses. On the other hand, we will be able to provide a lower
bound on the gaugino mass parameter $\Lambda_G$ by considering the
negative results of direct searches at colliders. We present results
for $M_{mess}= 10^{6}$~GeV, $10^{10}$~GeV and $10^{14}$~GeV corresponding to
low, medium and large messenger masses.
Any points that do not break electroweak
symmetry correctly, have tachyonic sparticles in their spectrum,
have couplings that become non-perturbative before the GUT scale,
have a scalar Higgs potential unbounded from below or have a
spectrum that violates the direct search limits detailed below are
excluded from our results. There are also points in our scans where
\texttt{SoftSUSY} has not numerically converged, due to the large
hierarchies between the scalars and gaugino masses. While we
therefore cannot present results on these regions they are not
necessarily theoretically excluded.

\begin{figure}
\begin{center}
\vspace*{-0.6cm}
\subfigure[]{
\includegraphics[bb= 142 75 500 400,clip,width=6.5cm]{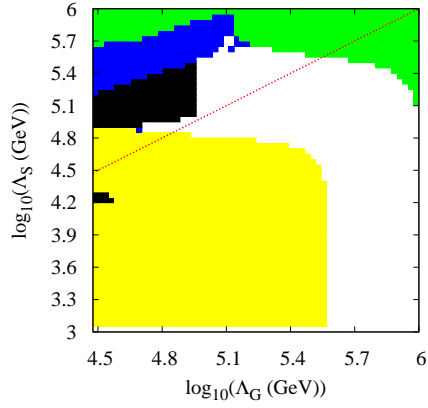}
}
\end{center}
\vspace*{-1.2cm}
\begin{center}
\subfigure[]{
\includegraphics[bb= 142 75 500 400,clip,width=6.5cm]{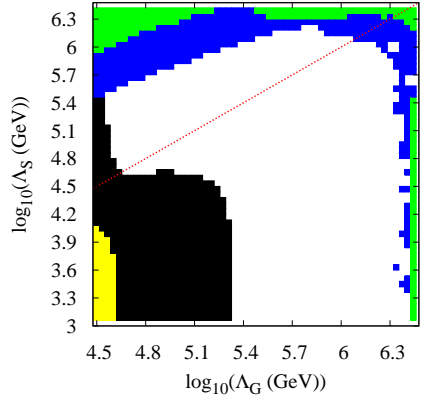}
}
\end{center}
\vspace*{-1.2cm}
\begin{center}
\subfigure[]{
\includegraphics[bb= 142 75 500 400,clip,width=6.5cm]{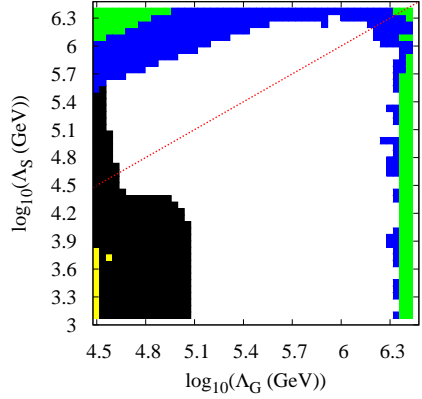}
}
\end{center}
\begin{center}
\vspace*{-0.5cm}
\caption{ Plots showing the constraints on parameter space for
(a) $M_{Mess} = 10^{6}$~GeV, (b) $M_{Mess} = 10^{10}$~GeV and (c) $M_{mess}= 10^{14}$~GeV.
Yellow (pale grey) means the point is excluded by the presence of tachyons in the spectrum, while the
black region falls foul of the direct search limits detailed in the text.
In the blue (dark grey) region \texttt{SoftSUSY} has not converged and in the green (light grey) region
a coupling reaches a Landau pole during RG evolution.
The red dotted line indicates the ordinary gauge mediation scenario where $\Lambda_G=\Lambda_S$.}

\label{fig:reasons}
\end{center}
\end{figure}

We have also applied constraints from direct searches at the Tevatron and
LEP, adapted from \cite{Amsler:2008zzb,Carpenter:2008he}. There is
an absolute lower limit of $45$~GeV on the chargino mass from LEP. If
the NLSP is the neutralino (with the LSP in all models of gauge mediation being, of course, the gravitino)
and the chargino-neutralino splitting is
less than $m_{\pi^+}$, limits from searches for long-lived
charged particles imply a lower bound of $206$~GeV for the chargino
mass. A promptly decaying chargino with chargino-neutralino
splitting greater than 3~GeV is required to be heavier than $229$~GeV,
while a non-promptly decaying chargino must be heavier than
$102.7$~GeV.

The lower limit on the gluino mass is $51$~GeV for non-gluino NLSP.
A promptly decaying gluino NLSP is required to be heavier than
$315$~GeV, while non-promptly decaying gluinos must be heavier than
$270$~GeV. The lightest squarks have a lower bound of $92$~GeV, while
the sneutrino must be heavier than $43$~GeV. For slepton NLSP, we
apply bounds of $68$~GeV for the stau and $85$~GeV for the selectron
and smuon when $M_{mess}=1\times10^6$~GeV. Searches at the OPAL detector~\cite{Abbiendi:2003yd}
place limits on slowly (i.e. the decay happens outside the detector) decaying NLSP sleptons
of $98$~GeV. For higher messenger masses,
$M_{mess}=10^{10,14}$~GeV, the stau NLSP is automatically
sufficiently slowly decaying. For $M_{mess}=10^{6}$~GeV this is not
guaranteed but the bound does not exclude any additional parameter
space anyway.
 For non-slepton NLSP
the selectron and smuon masses are required to be greater than $100$~GeV and
$95$~GeV respectively, while the stau mass must be greater
than $90$~GeV.
Finally, the absolute lower limit on the neutralino mass is $46$~GeV.
The promptly decaying neutralino NLSP has a lower bound of $125$~GeV,
while the non-prompt decay has a lower bound of $46$~GeV. In principle in General Gauge Mediation the neutralino
could be massless~\cite{Dreiner:2009ic}. This requires non-universal gaugino masses, and thus in our model with degenerate $\Lambda_G$ does not arise.

The direct search constraints lead to a lower bound on $\Lambda_G$ of $58$~TeV
for $M_{mess}=10^6$~GeV and $38$~TeV for $M_{mess}= 10^{10,14}$~GeV,
in agreement with \cite{Carpenter:2008he}. The region of low $\Lambda_G$ and $\Lambda_S$
is ruled out by tachyonic scalars and direct search limits. For larger values of $\Lambda_G$ the
contribution of the gaugino masses can increase the scalar masses during
running from $M_{mess}$ to low energies
by a sufficient amount so as to
result in phenomenologically viable scalar masses. It is this effect which is
responsible for the large region at very low $\Lambda_S$. Large values of $\Lambda_S$, and
hence large $\tan\beta$ can lead to non-perturbativity before the GUT scale.
The large masses involved also make it difficult for the numerical routines
in \texttt{SoftSUSY} to converge. These effects rule out the regions at the
regions at high $\Lambda_S$.
This is what cuts off the allowed region in parameter space in the top-left corner
where $\tan\beta$ becomes very large.
Most of the parameter space for split SUSY spectra, $\Lambda_G\ll\Lambda_S$, is not viable.
However, it is important to stress that the remaining region (close to this boundary) still allows for
a mildly split SUSY with $\Lambda_G/\Lambda_S\sim 1/10$.
This is where the direct gauge mediation models considered
in Refs.~\cite{Izawa:1997gs,Kitano:2006xg,Csaki:2006wi,Abel:2007jx,Abel:2007nr,Abel:2008gv}
can live and is the location of the benchmark point
presented in Ref.~\cite{Abel:2007nr}.

The constraints described above are summarised in Fig.~\ref{fig:reasons} (a,b,c) which correspond to $M_{mess} = 1\times
10^{6,10,14}$~GeV respectively. Yellow regions have a tachyon in the spectrum. Black regions are in principle viable, but are ruled out by the direct searches. The blue region indicates where \texttt{SoftSUSY} did not converge, and in the green area a coupling encounters a Landau pole.
Not surprisingly lack of convergence indicates that one is in the proximity of a Landau pole.

\begin{figure}
\vspace*{-0.6cm}
\begin{center}
\subfigure[]{
\includegraphics[bb= 142 75 500 400,clip,width=6.5cm]{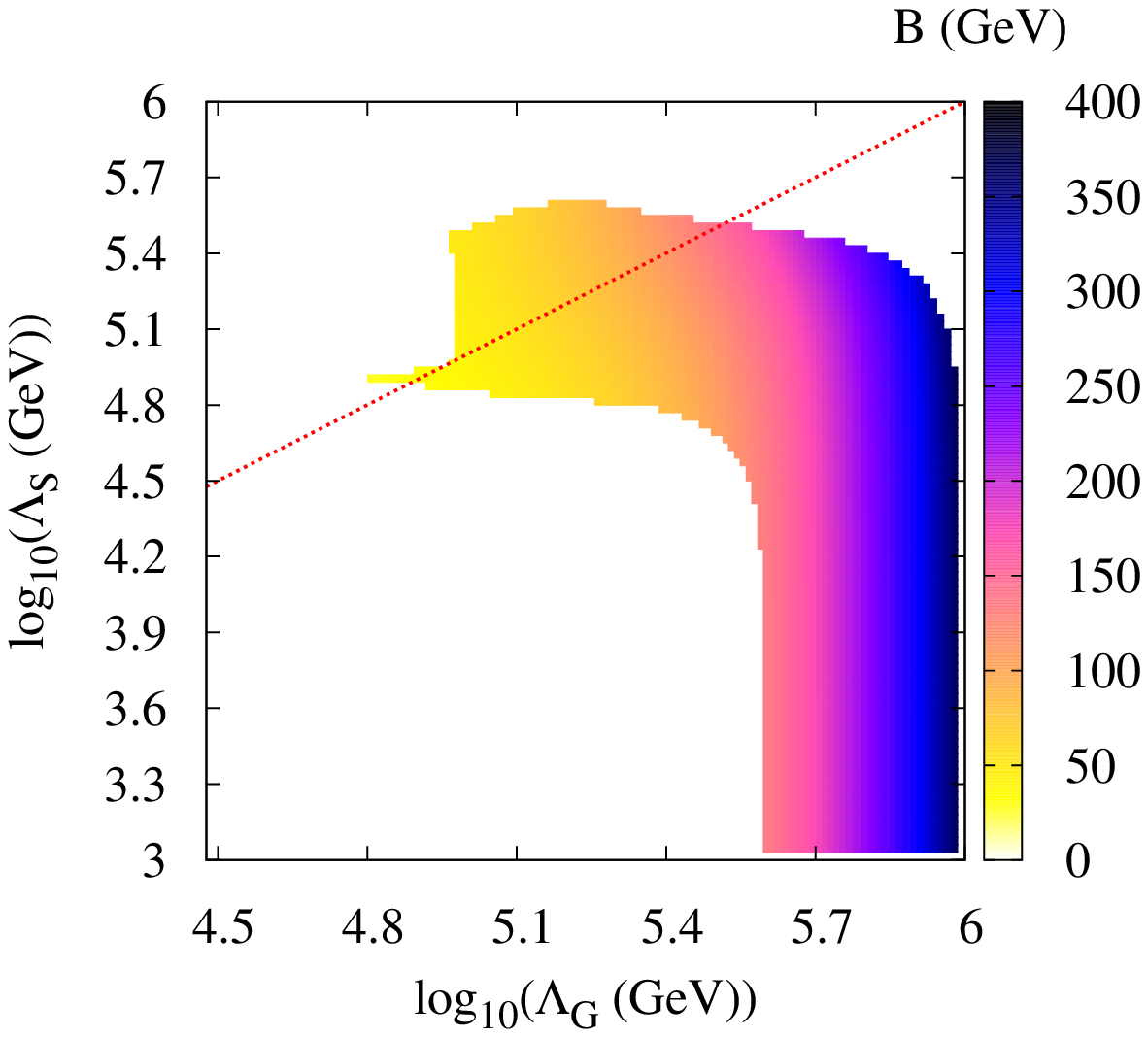}
}
\hspace*{1cm}
\subfigure[]{
\includegraphics[bb= 142 75 500 400,clip,width=6.5cm]{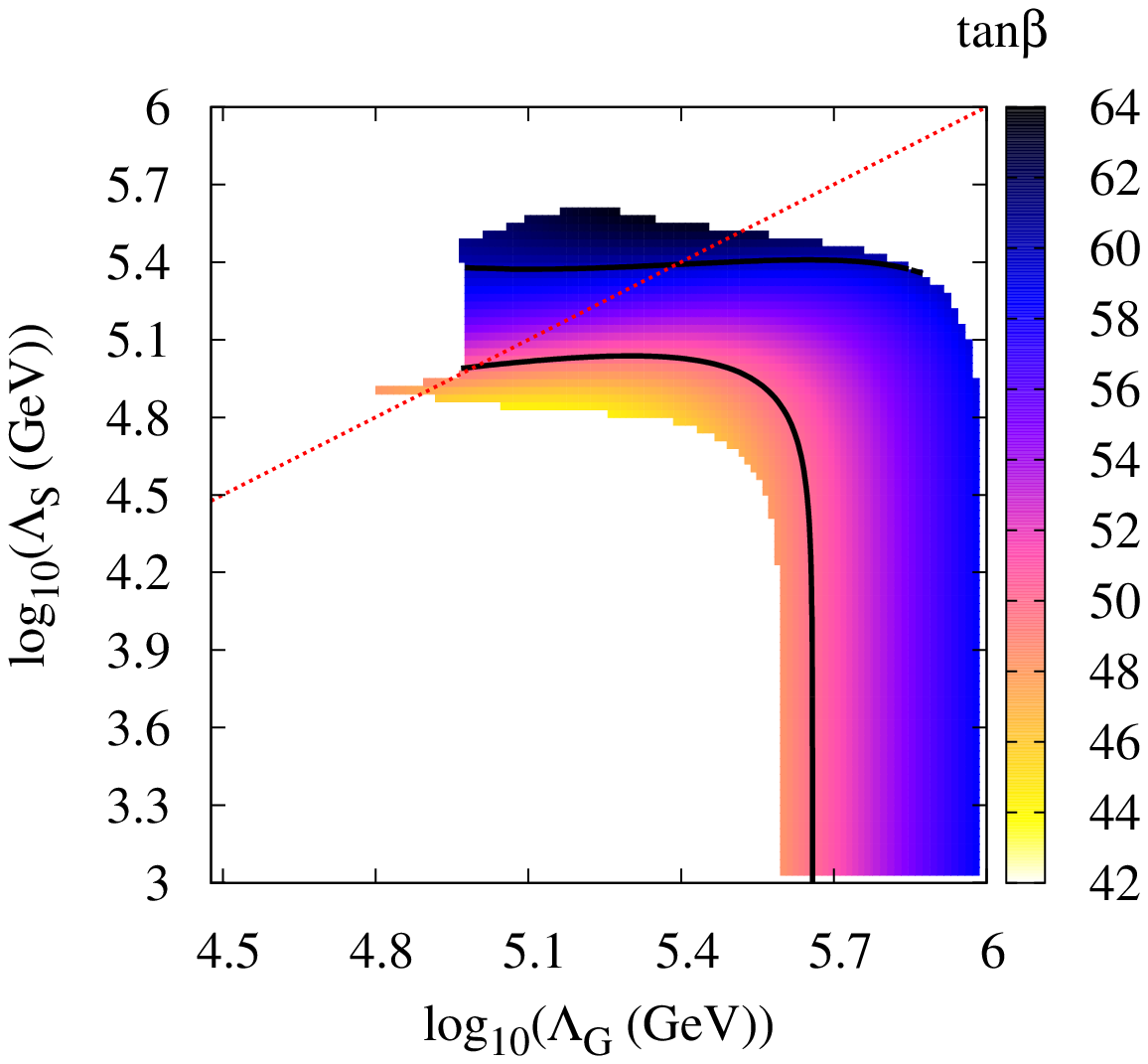}
}
\\
\vspace*{-0.7cm}
\subfigure[]{
\includegraphics[bb= 142 75 500 400,clip,width=6.5cm]{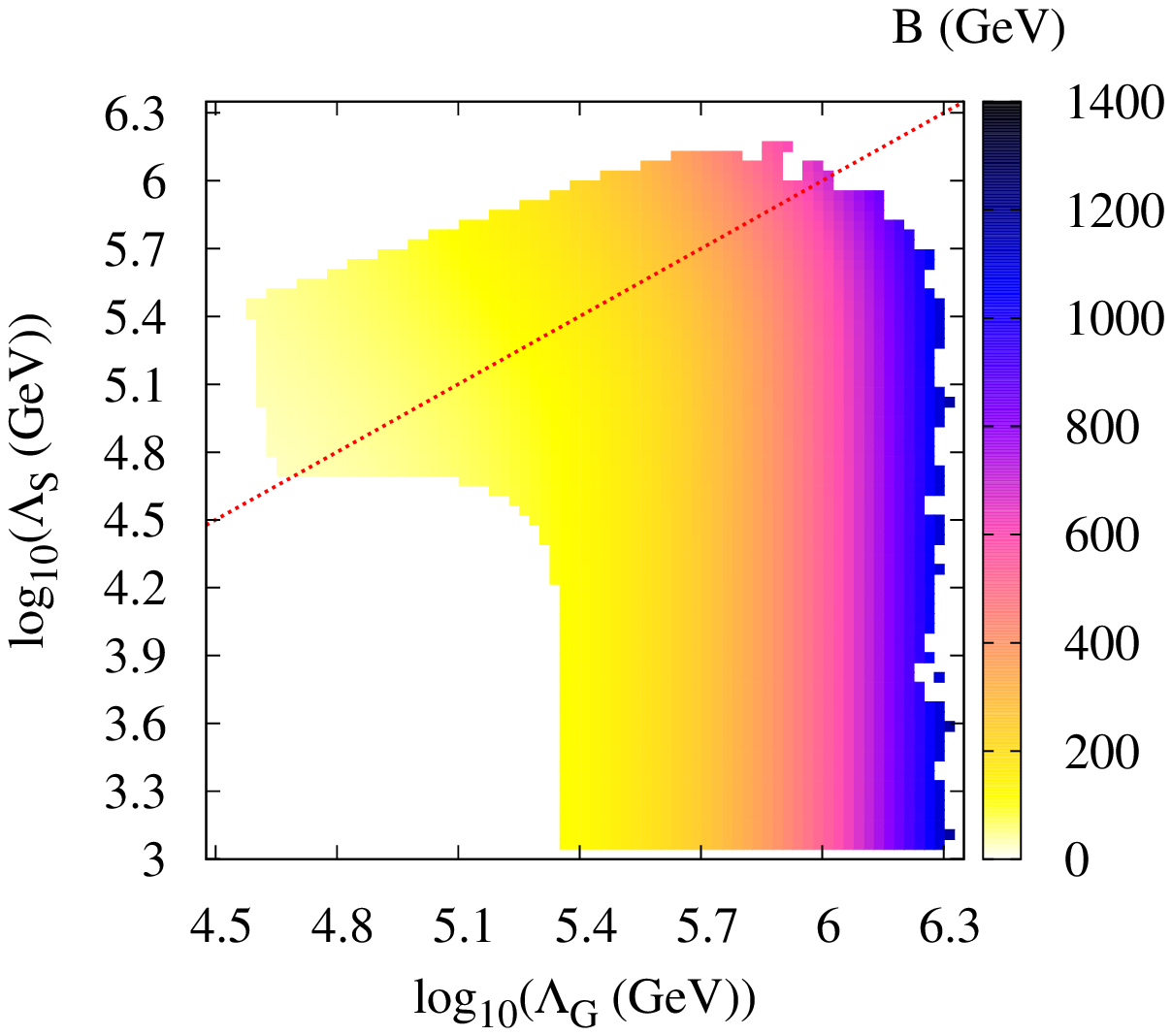}
}
\hspace*{1cm}
\subfigure[]{
\includegraphics[bb= 142 75 500 400,clip,width=6.5cm]{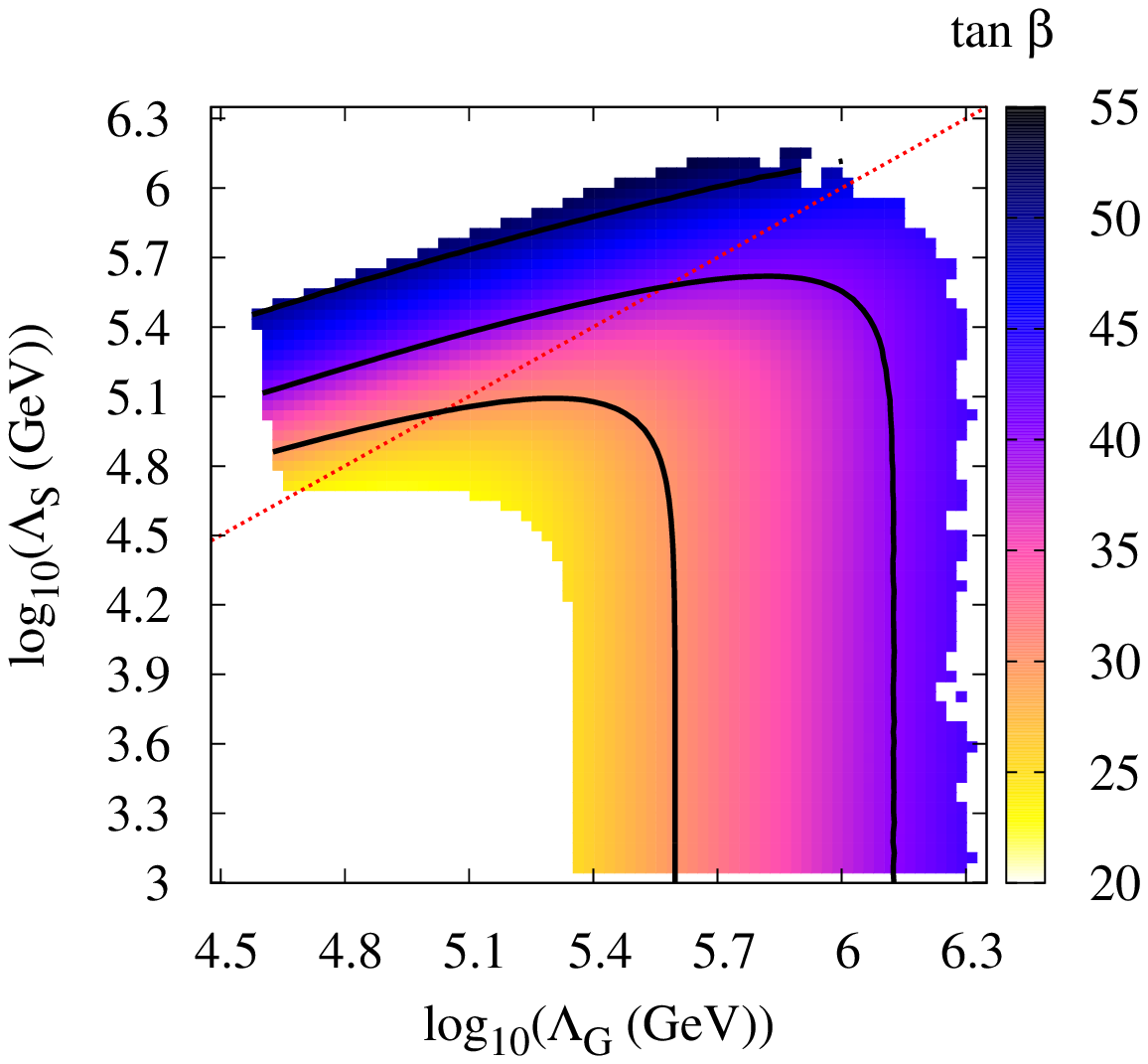}
}
\\
\vspace*{-0.7cm}
\subfigure[]{
\includegraphics[bb= 142 75 500 400,clip,width=6.5cm]{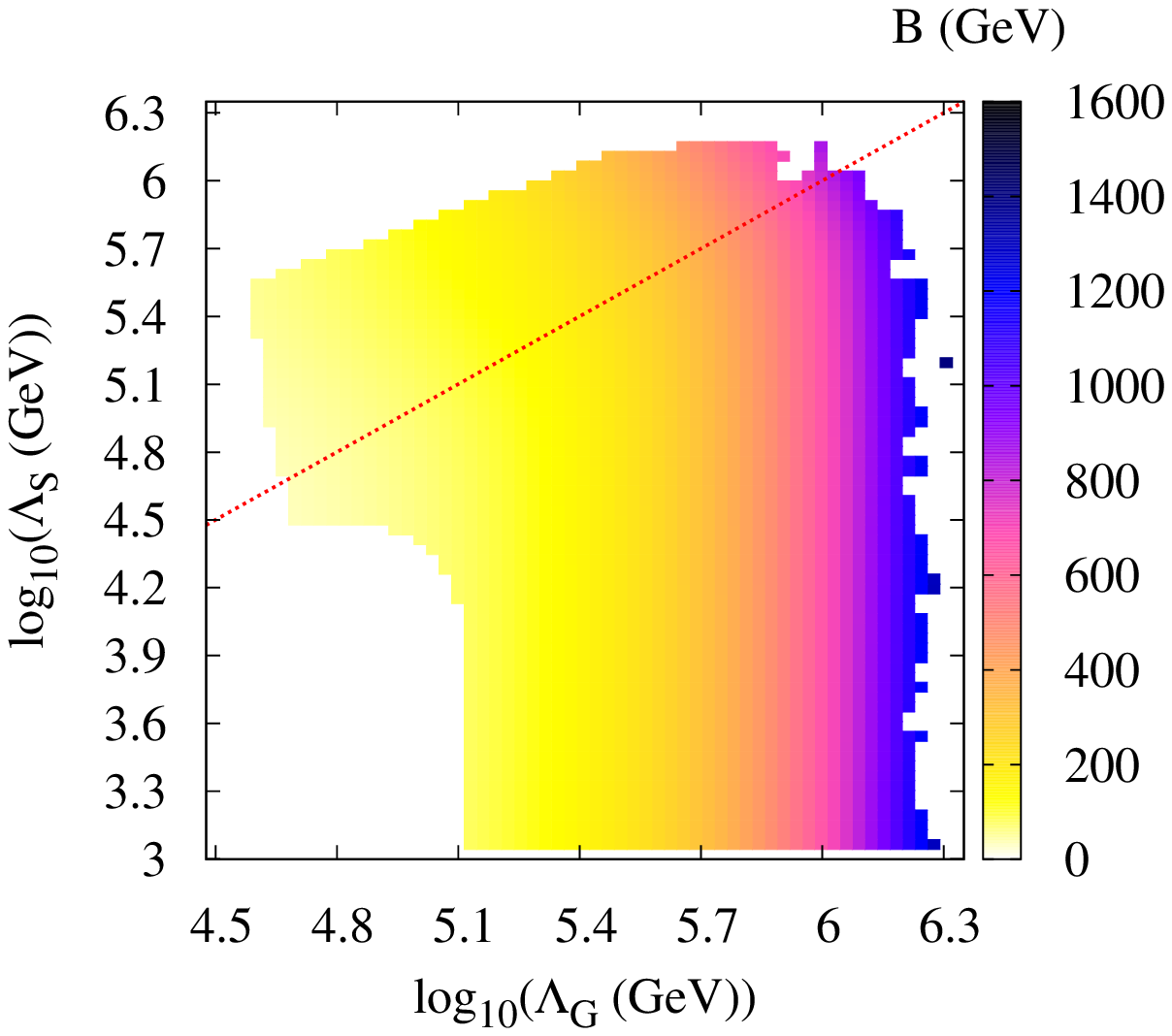}
}
\hspace*{1cm}
\subfigure[]{
\includegraphics[bb= 142 75 500 400,clip,width=6.5cm]{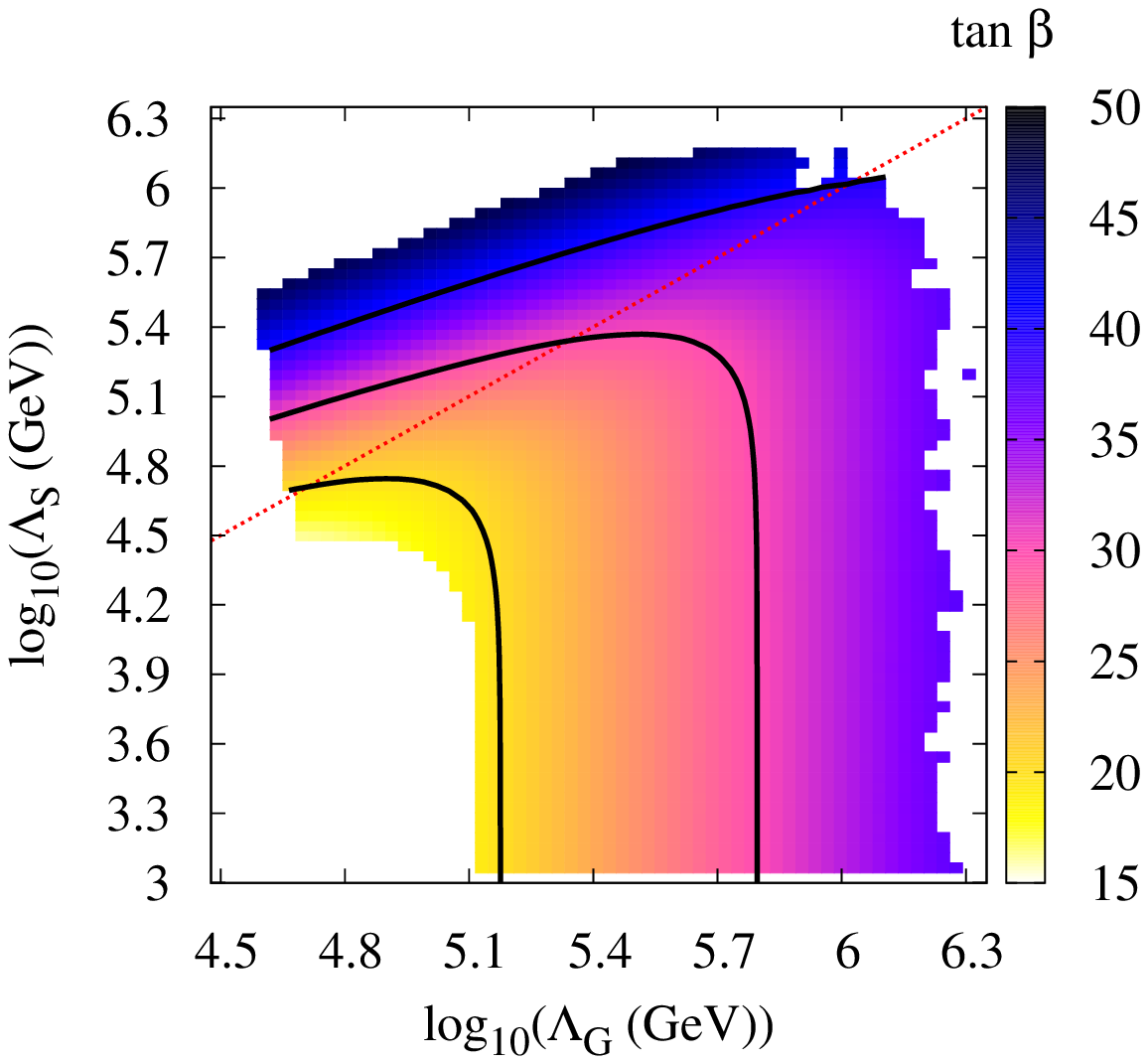}
}
\caption{The low energy values of the $B$ parameter for
(a) $M_{mess}=10^6$~GeV, (c) $M_{mess}=10^{10}$~GeV and (e) $M_{mess}=10^{14}$~GeV. Figure \ref{fig:lowB} (b,d,f) shows the values of $\tan\beta$ obtained from the electroweak symmetry breaking conditions for (b) $M_{mess}=10^6$~GeV, (d) $M_{mess}=10^{10}$~GeV and (f) $M_{mess}=10^{14}$~GeV along with contours of $\tan\beta =20,30,40,50$ and $60$.
Note that the scales of the colour
coding are different for each plot. The red dotted line indicates the
ordinary minimal gauge mediation scenario where $\Lambda_G=\Lambda_S$.}
\label{fig:lowB}
\end{center}
\end{figure}

Let us now turn to some specific properties of our scenario.
As already explained we take $B_\mu=0$ at the high (messenger) scale.
As expounded in some detail in Section \ref{sec:appB}, it
is then radiatively generated and its value at the low scale is roughly proportional
to the gaugino mass parameter $\Lambda_G$. This dependence on $\Lambda_G$ can be clearly seen
in Figure \ref{fig:lowB} (a,c,e) where we show the low energy values of $B=B_{\mu}/\mu$
for $M_{mess}=10^{6,10,14}\,{\rm GeV}$ in the $\Lambda_G$-$\Lambda_S$ plane.
$B$ does not strongly depend
on the scalar mass parameter $\Lambda_S$.
In the region of light scalars and gauginos, $B$ is a few hundred GeV and increases
with $\Lambda_G$.
We also show the red dotted line corresponding to minimal gauge mediation $\Lambda_G=\Lambda_S$ which
was already investigated in Ref.~\cite{Rattazzi:1996fb}.

The fact that $\tan\beta$ is a determined parameter rather than a free input is
where our approach markedly differs from previous studies such as that in~\cite{Carpenter:2008he,Rajaraman:2009ga}.
Figure \ref{fig:lowB} shows the values of $\tan\beta$ obtained for
(b) $M_{mess} = 10^6$, (d) $10^{10}$~GeV and (f) $10^{14}$~GeV, along
with contour lines of $\tan\beta = 20,30,40,50$ and $60$.
In general we expect large $\tan\beta$ since $B_\mu$ is small.
Since the low energy value of $B_\mu$ increases with $\Lambda_G$ we therefore would expect $\tan\beta$ to decrease as
$\Lambda_G$ increases. However, this holds only when when the running of the Higgs masses
is dominated by $\Lambda_S$ and thus we observe this behavior only in the
top parts of Figure~\ref{fig:lowB}(b,d,f).
As $B_\mu$ will be smaller for low values of $M_{mess}$, $\tan\beta$ will
be correspondingly higher for $M_{mess}=10^6$~GeV than
for $M_{mess}=10^{10}$~GeV or $10^{14}$~GeV.
For $M_{mess} = 10^{6}$~GeV the minimum
value of $\tan\beta$ is 43.9, for $M_{mess} = 10^{10}$~GeV
it is 22.2 and for $M_{mess}=10^{14}$~GeV the minimum value is 15.8. The maximum values of $\tan\beta$ are 63.7,
53.9 and 48.4 for $M_{mess}= 10^{6,10,14}$~GeV, respectively.

\begin{figure}
\vspace*{-0.6cm}
\begin{center}
\subfigure[]{
\includegraphics[bb= 142 75 500 400,clip,width=6.5cm]{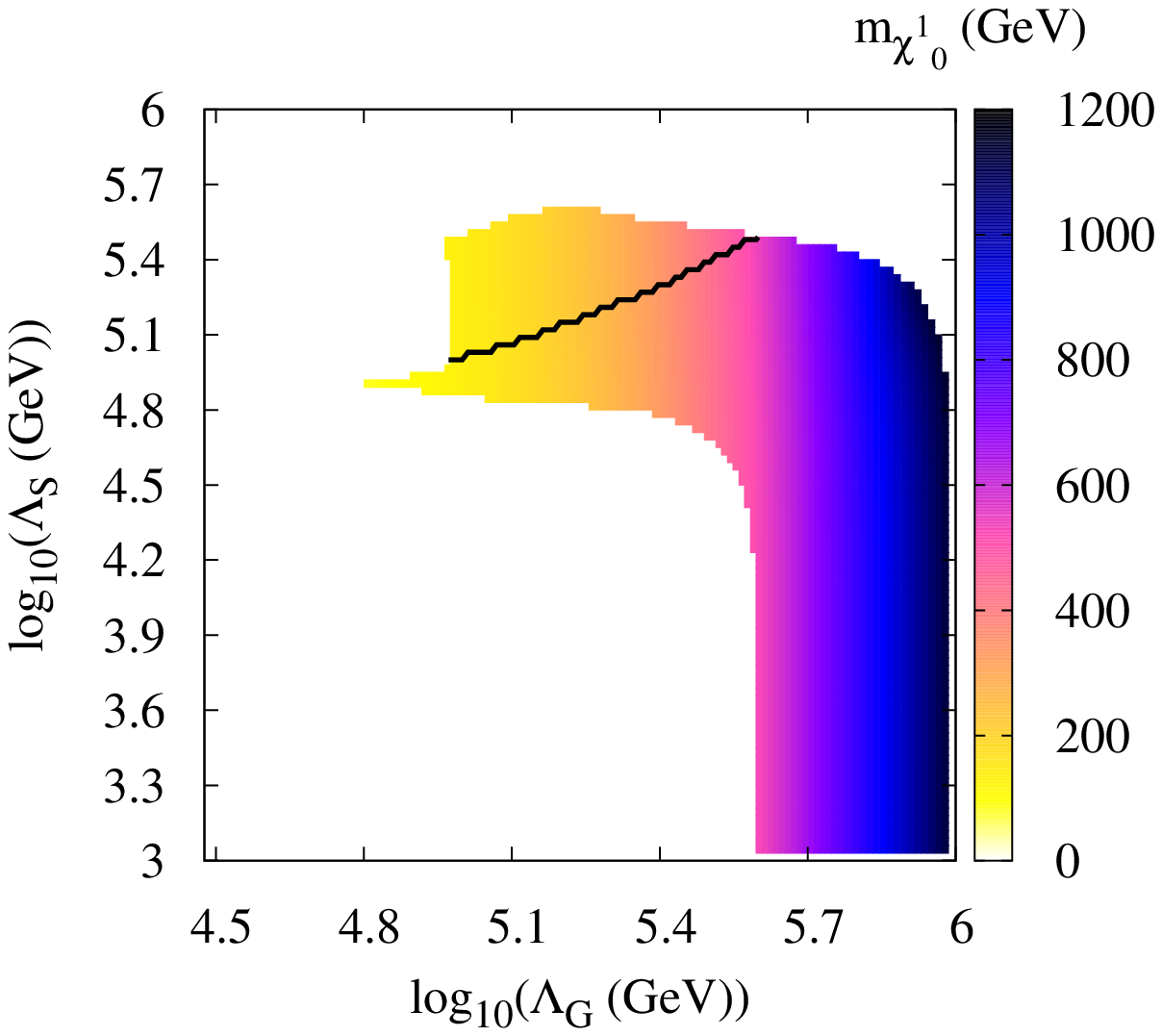}
}
\hspace*{1cm}
\subfigure[]{
\includegraphics[bb= 142 75 500 400,clip,width=6.5cm]{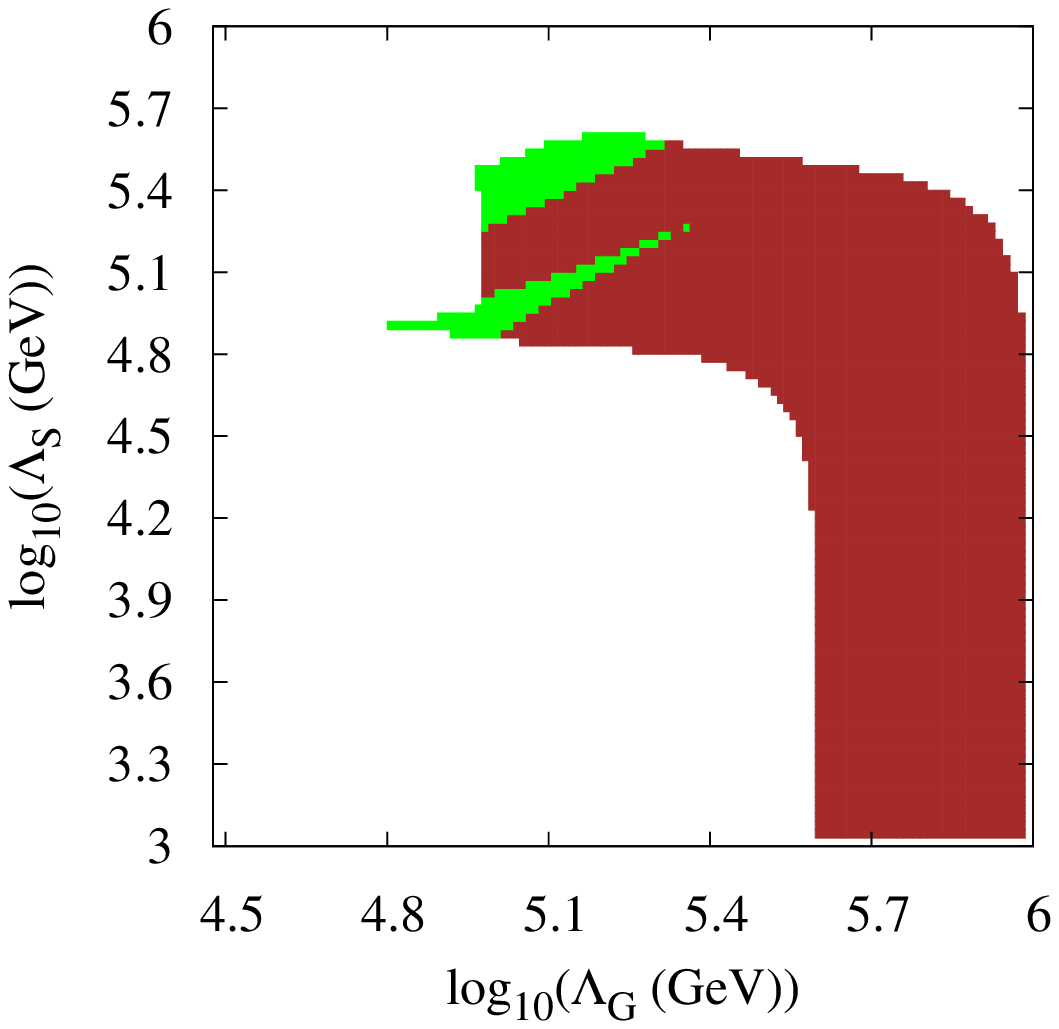}
}
\\
\vspace*{-0.7cm}
\subfigure[]{
\includegraphics[bb= 142 75 500 400,clip,width=6.5cm]{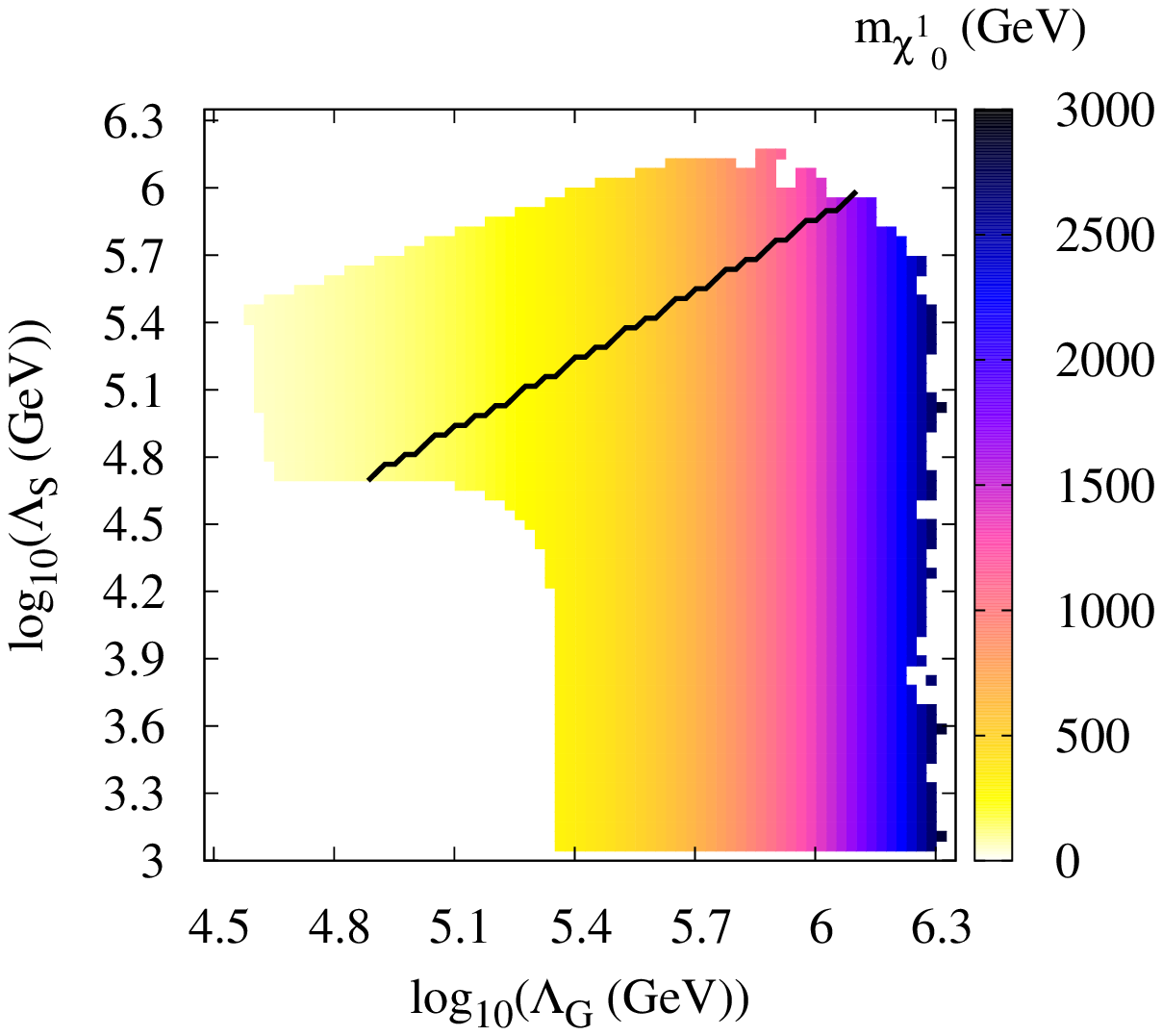}
}
\hspace*{1cm}
\subfigure[]{
\includegraphics[bb= 142 75 500 400,clip,width=6.5cm]{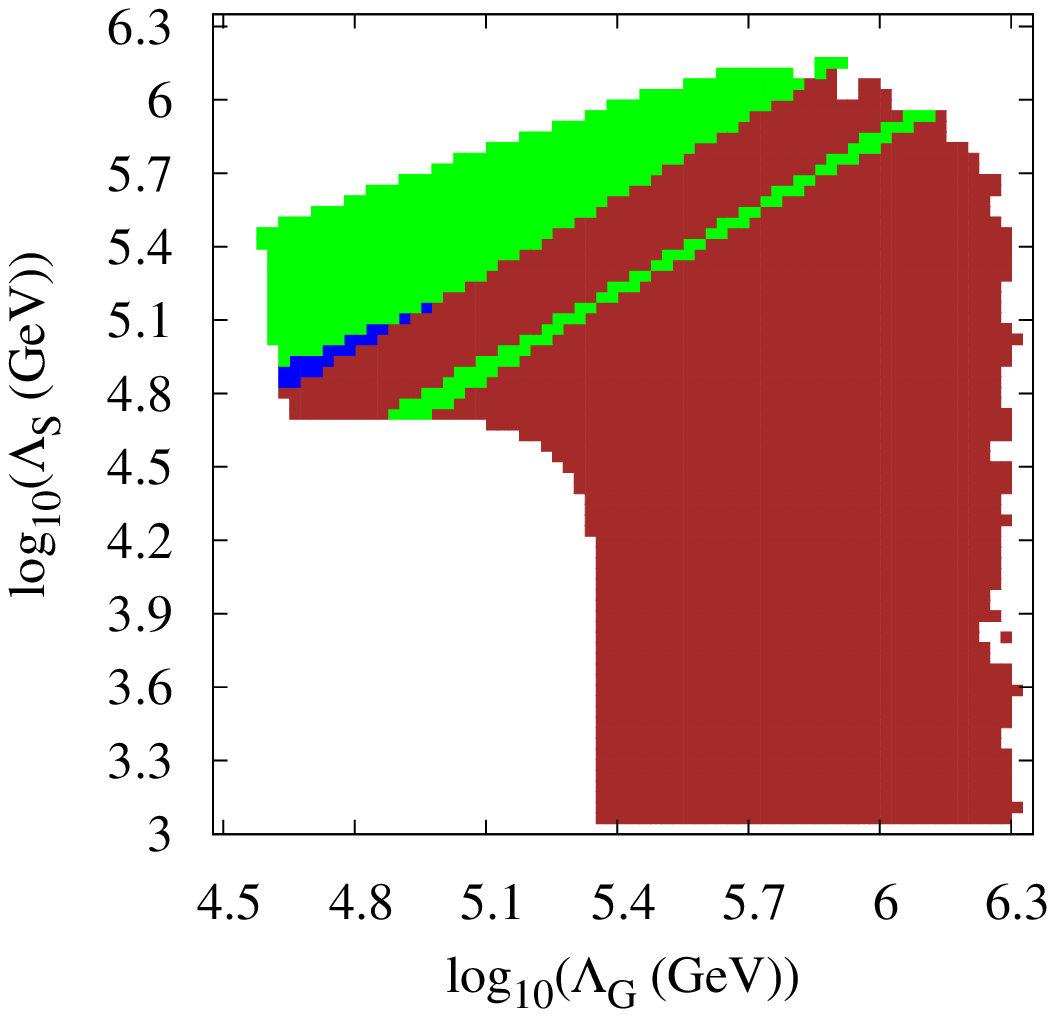}
}
\\
\vspace*{-0.7cm}
\subfigure[]{
\includegraphics[bb= 142 75 500 400,clip,width=6.5cm]{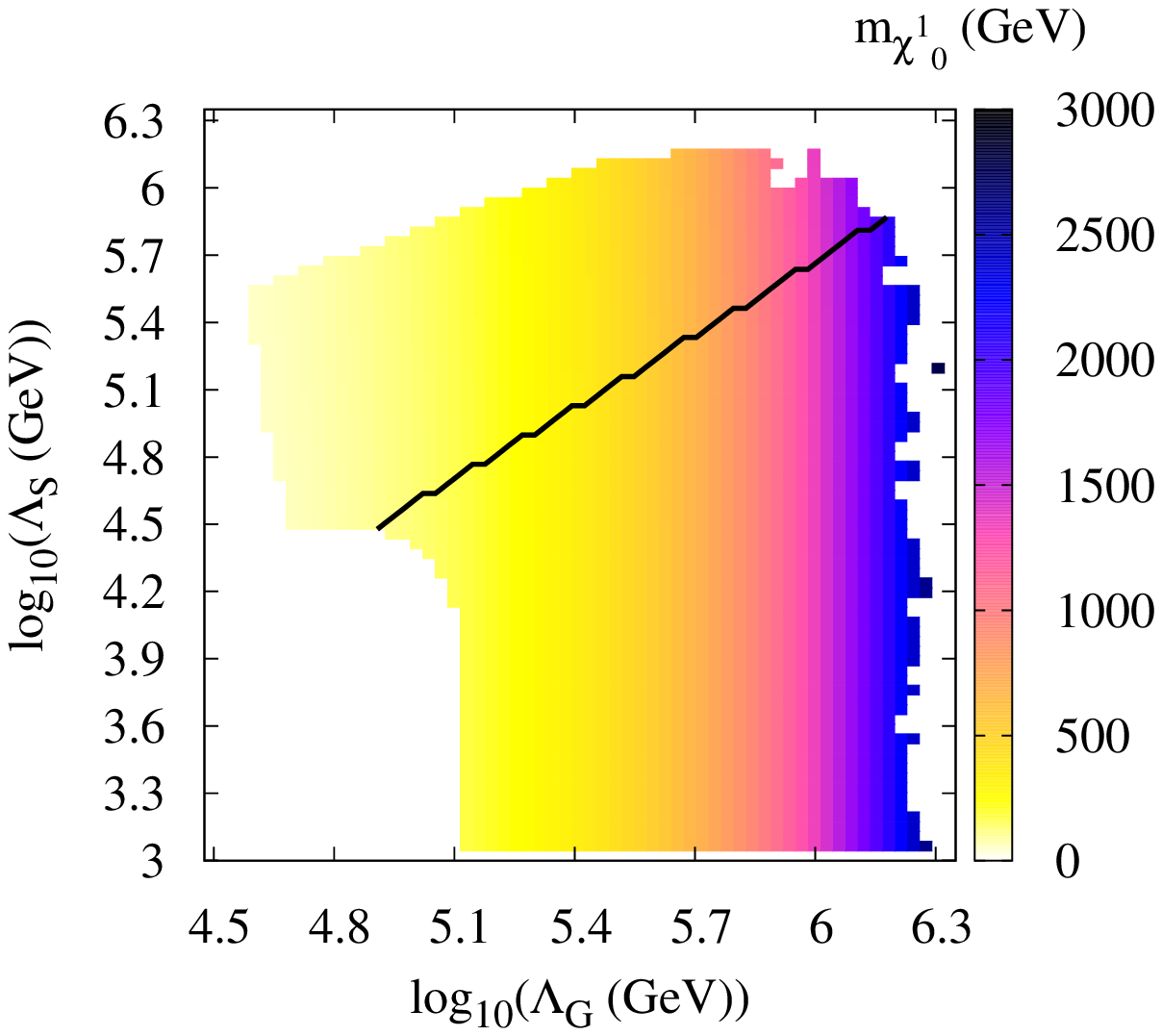}
}
\hspace*{1cm}
\subfigure[]{
\includegraphics[bb= 142 75 500 400,clip,width=6.5cm]{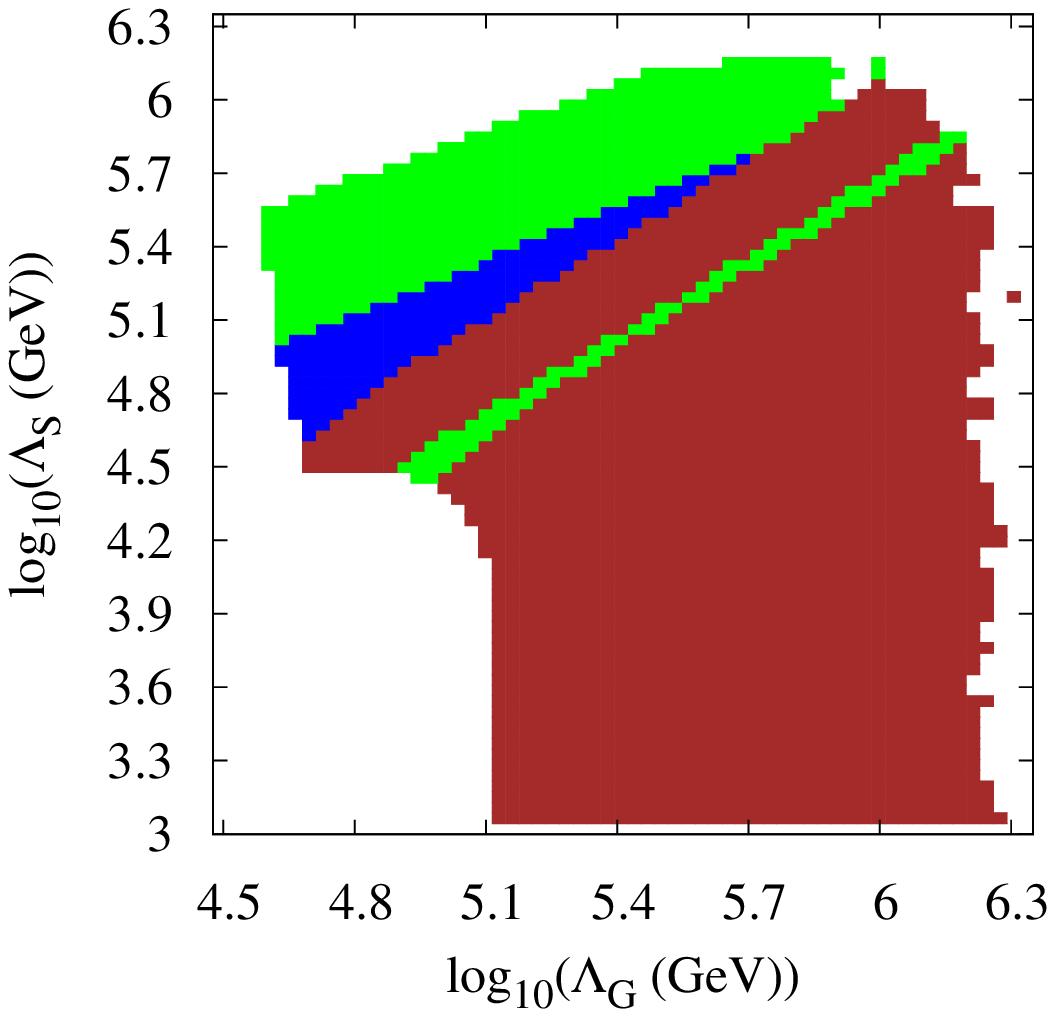}
}
\caption{Details of the spectrum for (a,b) $M_{mess}=10^6$, (c,d) $M_{mess}=10^{10}$ and (e,f) $M_{mess}=10^{14}$~GeV. Figure \ref{fig:spectrum} (a,c,e) show the lightest neutralino mass. Above the black line the NLSP is neutralino, below it is the
lightest slepton (usually the stau, sometimes the smuon). Figure \ref{fig:spectrum} (b,d,f) shows the NNLSP species. Green is neutralino, brown is a slepton and blue is the lightest chargino.}
\label{fig:spectrum}
\end{center}
\end{figure}

\begin{figure}
\vspace*{-0.6cm}
\begin{center}
\subfigure[]{
\includegraphics[bb= 142 75 500 400,clip,width=6.5cm]{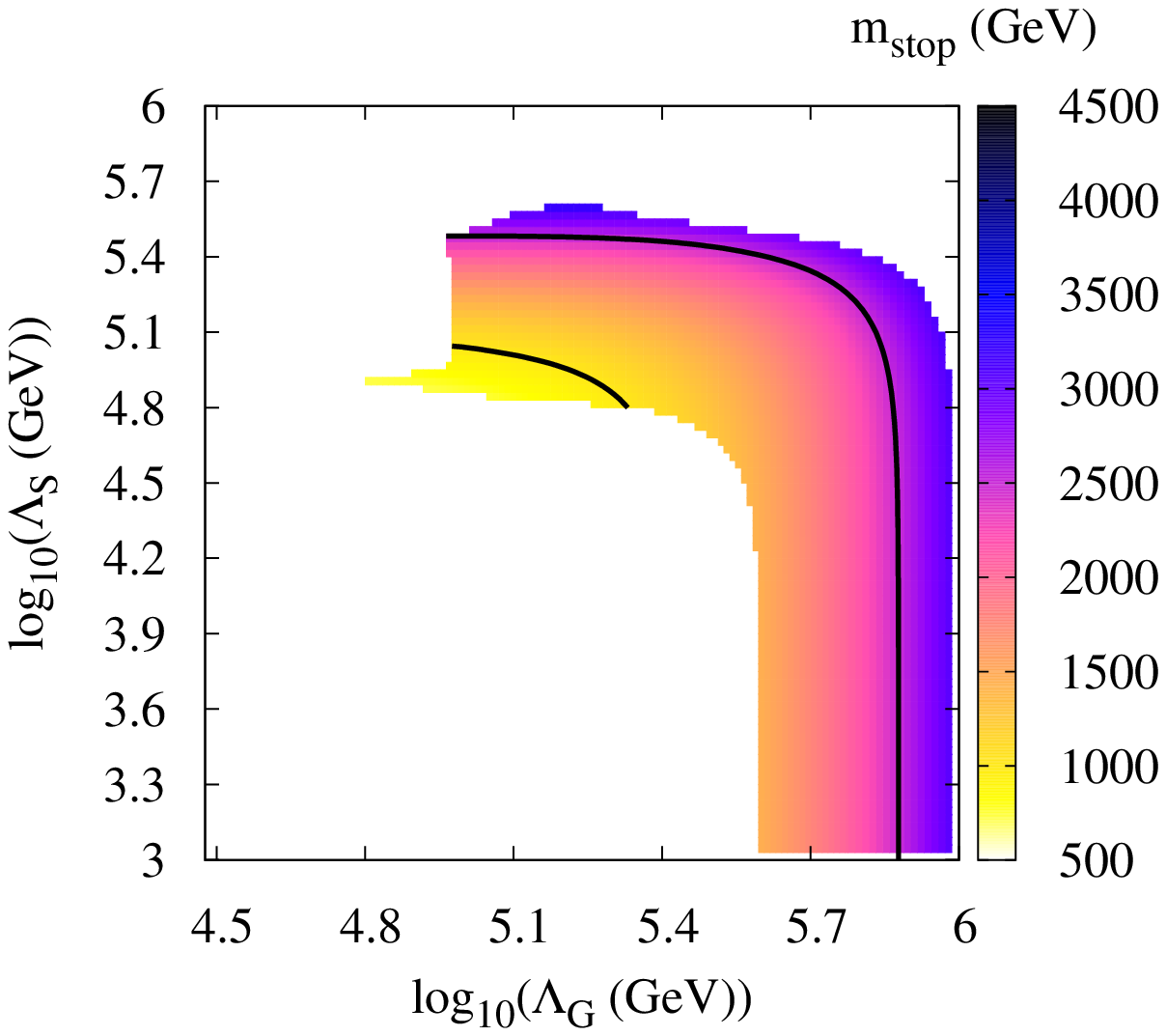}
}
\hspace*{1cm}
\subfigure[]{
\includegraphics[bb= 142 75 500 400,clip,width=6.5cm]{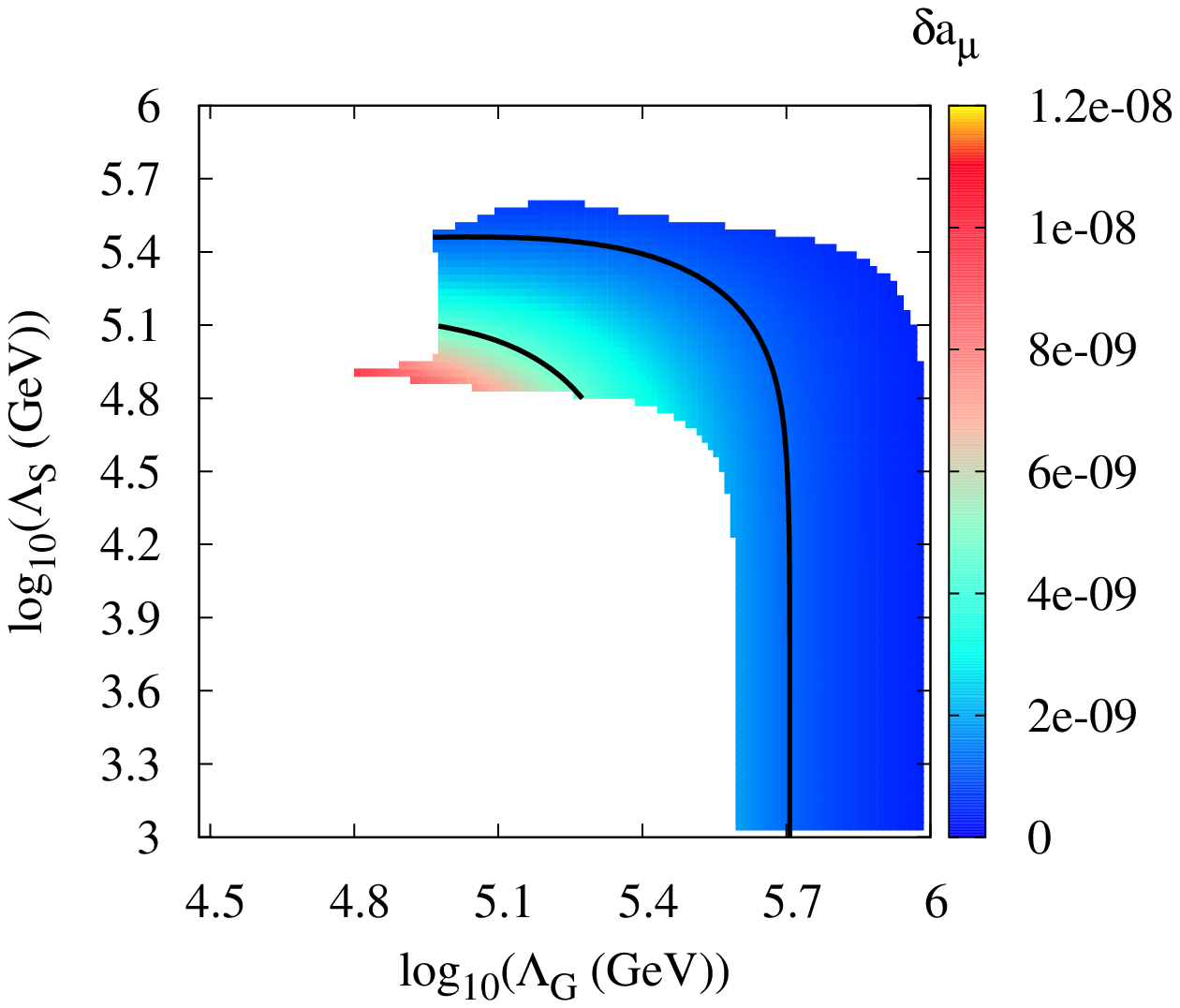}
}
\\
\vspace*{-0.7cm}
\subfigure[]{
\includegraphics[bb= 142 75 500 400,clip,width=6.5cm]{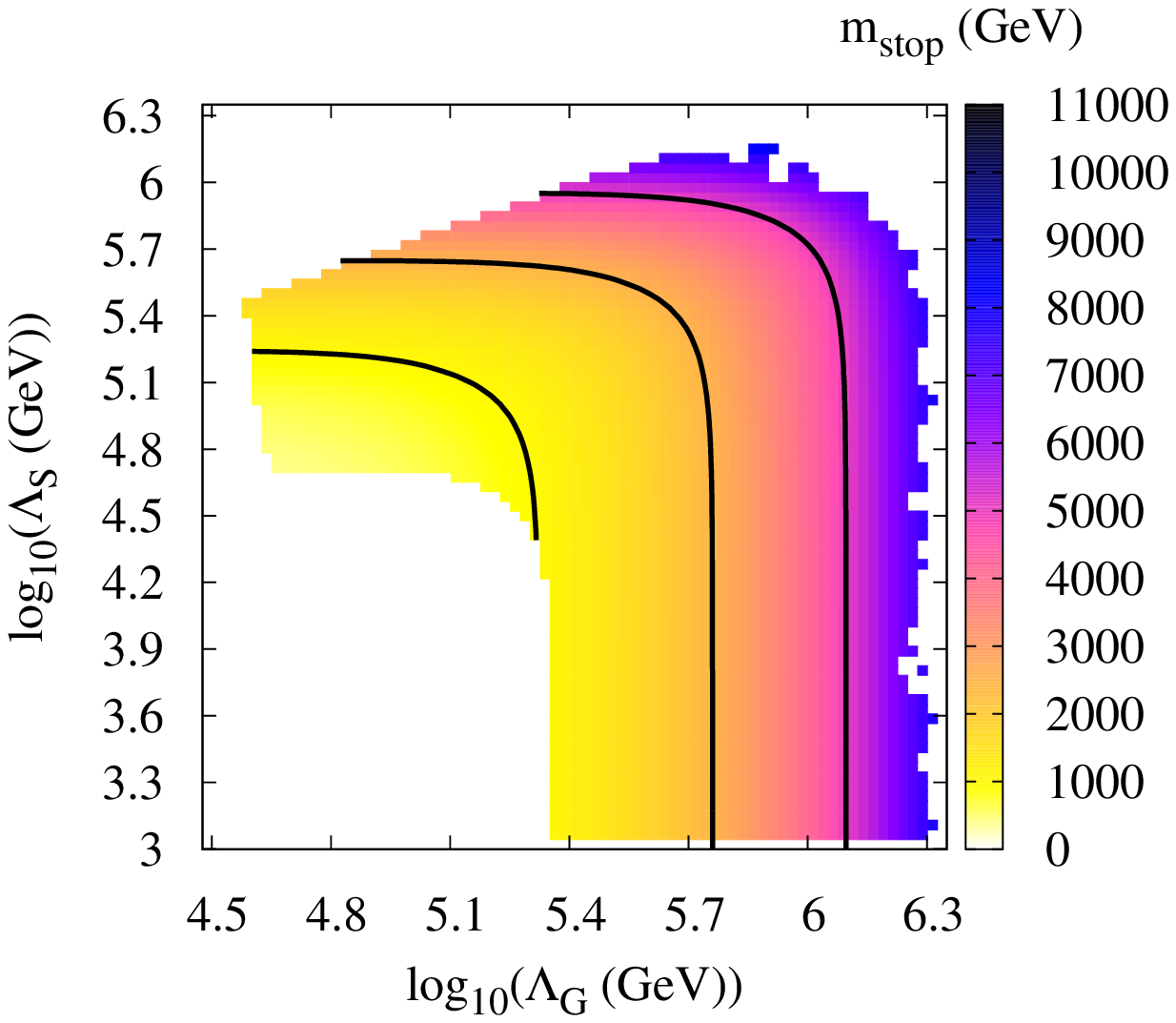}
}
\hspace*{1cm}
\subfigure[]{
\includegraphics[bb= 142 75 500 400,clip,width=6.5cm]{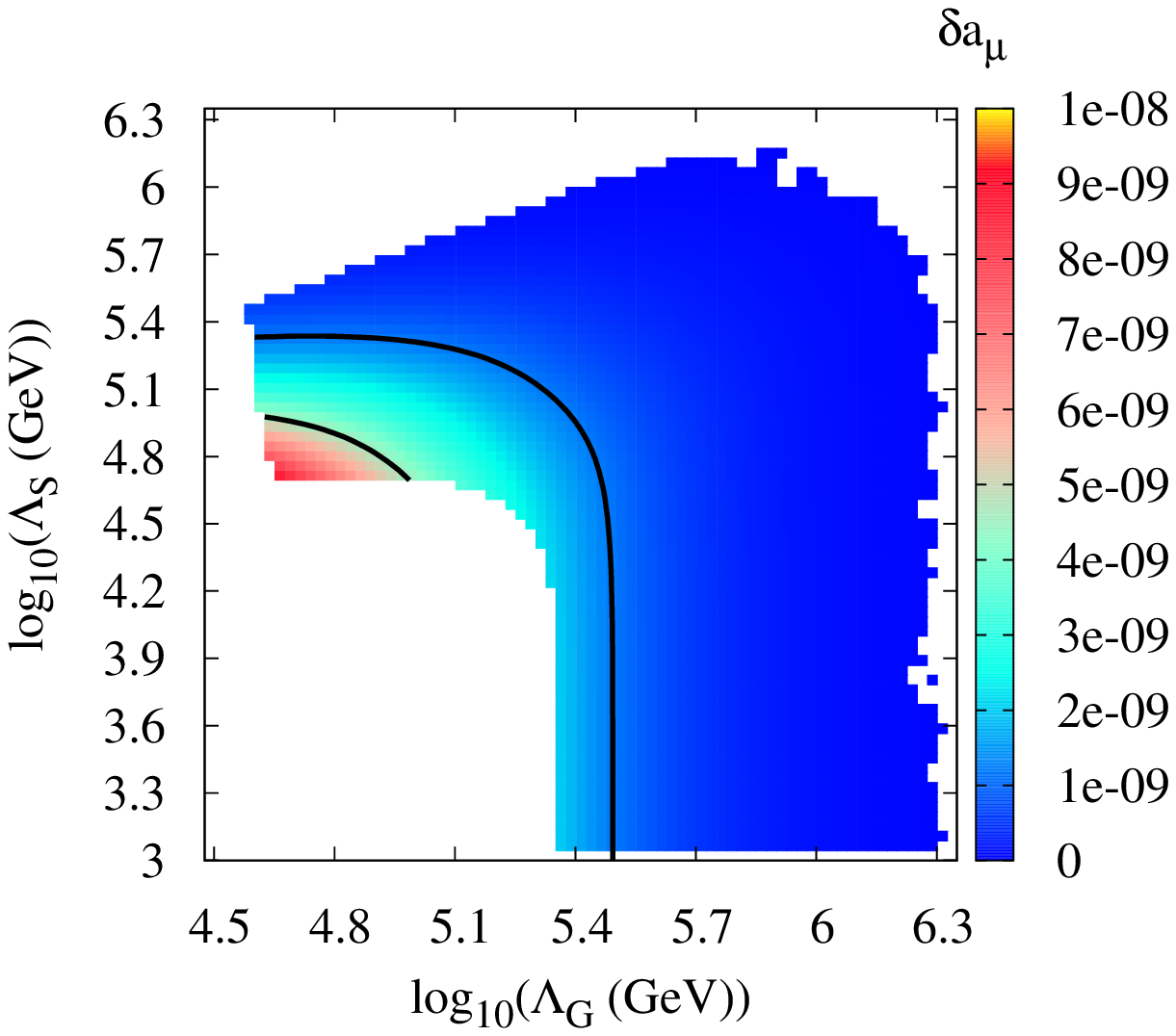}
}
\\
\vspace*{-0.7cm}
\subfigure[]{
\includegraphics[bb= 142 75 500 400,clip,width=6.5cm]{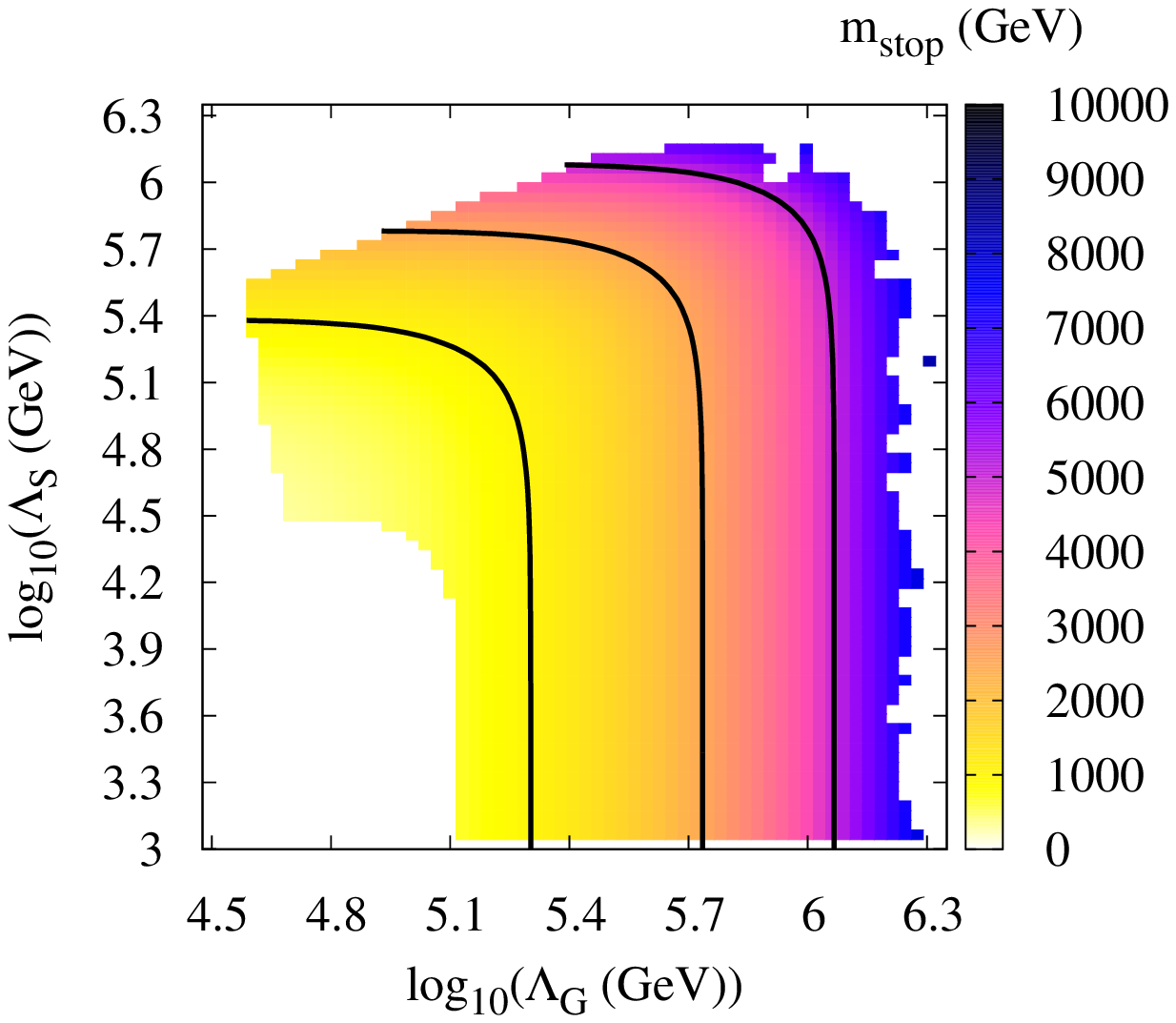}
}
\hspace*{1cm}
\subfigure[]{
\includegraphics[bb= 142 75 500 400,clip,width=6.5cm]{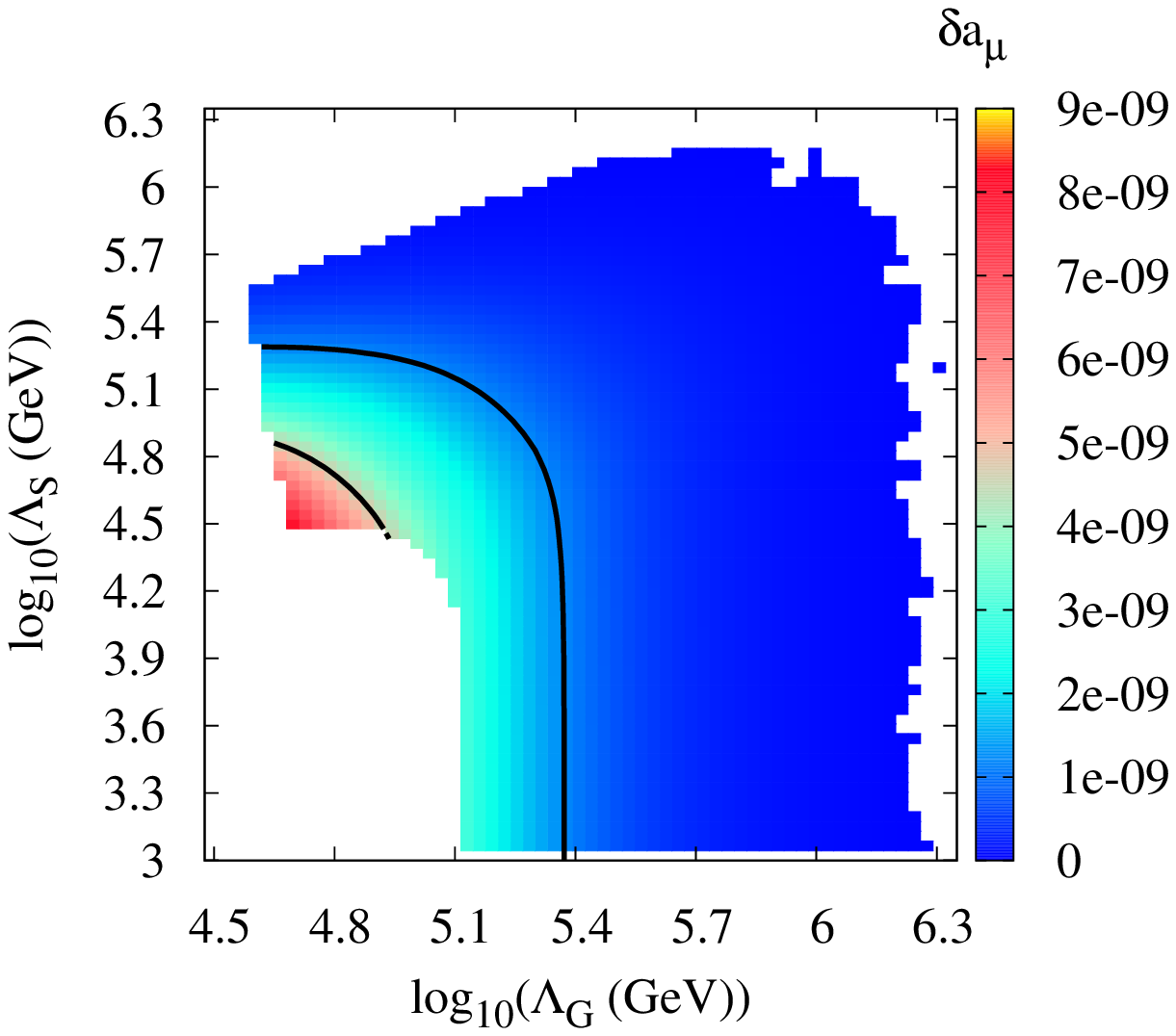}
}
\caption{Further phenomenological features for (a,b) $M_{mess}=10^6$, (c,d) $M_{mess}=10^{10}$ and (e,f) $M_{mess}=10^{14}$~GeV. Figure (a,c,e) show the mass of the lightest stop $m_{\tilde{t}_1}$, with contours of 1, 2.5 and 5~TeV. Figure (b,d,f) shows the values of the
anomalous magnetic moment of the muon $\delta a_{\mu}$ along with the $\pm1\sigma$ contours.}
\label{fig:g-2}
\end{center}
\end{figure}

Considering the spectrum, Figure \ref{fig:spectrum} (a,c,e) shows the mass of the lightest neutralino for  $M_{mess} = 10^{6,10,14}$ respectively.
The black contour line bisecting the plots diagonally indicates the nature of the NLSP: above the black line the NLSP is the lightest neutralino, while
below it the NLSP is the lightest slepton, the stau.
The stau NLSP could provide for a rather interesting signature. Since the NLSP decay into gravitinos can be quite small,
the stau NLSP
can behave like a metastable charged particle that can even decay outside the detector.
Such particles have rather interesting
phenomenological features as discussed, e.g., in Refs.~\cite{Raklev:2009mg,Feng:2004yi,DeRoeck:2005bw,Hamaguchi:2006vu}.
Unlike Ref.~\cite{Rajaraman:2009ga} we do not find any regions of sneutrino or chargino NLSP.

The collider phenomenology of gauge mediated theories is determined
to a large extent by the NLSP and NNLSP.
Accordingly Figures \ref{fig:spectrum} (b,d,f) show the species of NNLSP according
to the following scheme: green is the second lightest neutralino,
brown is a slepton (either the stau or the smuon) and blue is the lightest chargino.
For neutralino NLSP there is a significant proportion of parameter space which also
has neutralino NNLSP. In this case the three body decay $\chi_2^0 \to \chi_1^0 l^+ l^-$ is
most likely to occur. In the region of low scalar mass with stau NLSP and slepton NNLSP, large $\tan\beta$ implies that the splitting between the NLSP and the heavier $\tilde{e}$ and $\tilde{\mu}$ can be reasonably big, leading to 3 body slepton decay into a stau, $\tilde{l}^{\pm} \to \tilde{\tau}^{\pm}\tau^{\mp} l^{-}$, with  generally larger production cross sections for sleptons than for squarks.
Figure \ref{fig:g-2} (a,c,e) shows the mass of the lightest squark, the stop,
along with mass contour lines at 1, 2.5 and 5~TeV for $M_{mess}=10^{6}$~GeV, $10^{10}$~GeV and, $10^{14}$~GeV respectively.

In the MSSM, decay processes that are naively suppressed by loop factors can become enhanced by factors of $\tan\beta$. This applies in particular to flavor changing neutral current (FCNC) processes. Precision measurements in the B-physics sector therefore already present strong constraints on a light spectrum of superparticles.
Accordingly, we have calculated in our scans several low energy constraints which we
have applied to the parameter space of our effective model to bound the
possible values of $\Lambda_G/\Lambda_S$.
We show the observables we have used in Table~\ref{tab:obs}, along with their experimentally determined values.
We include the anomalous magnetic moment of the muon, calculated to one-loop
by \texttt{micrOMEGAS}~\cite{Belanger:2008sj}, with extra code to calculate the logarithmic piece of the QED 2-loop calculation,
the 2-loop stop-higgs, the chargino-stop/bottom contributions~\cite{Heinemeyer:2003dq,Heinemeyer:2004yq}, and the $\tan^2\beta$ enhanced two-loop contribution due to the shift between the muon mass and Yukawa coupling~\cite{Marchetti:2008hw}.
 As an example of a low energy
observable we have plotted the deviation of the anomalous magnetic moment
of the muon from its Standard Model prediction, $\delta a_{\mu}=(g-2)_{\mu}-(g-2)_{\mu}|_{SM}$
in Figure~\ref{fig:g-2}~(b,d,f), along with the $\pm1\sigma$ contours.
The majority of our observables are from the $B$-physics sector, as it is this area that combines sensitivity to the MSSM spectrum with experimental constraints to the greatest extent.
We include the rare branching ratios $BR(B\to X_s \gamma)$, $BR(B_s \to \mu^+ \mu^-)$, $BR(B\to\tau\nu)$ and  $BR(B \to D \tau \nu)$. We use SuperIso~\cite{Mahmoudi:2007vz,Mahmoudi:2008tp} to calculate the the isospin asymmetry
\begin{equation}
\Delta_{0-} = \frac{\Gamma(\overline{B}^0 \to \overline{K}^{*0}\gamma)-\Gamma(B^{\pm} \to K^{*\pm}\gamma)}
{\Gamma(\overline{B}^0 \to \overline{K}^{*0}\gamma)+\Gamma(B^{\pm} \to K^{*\pm}\gamma)}
\end{equation}
of the decay $B\to K^{*}\gamma$.
Finally, we also include the supersymmetric contribution to
the mass splitting of the $B_s$ meson and the supersymmetric contribution $R_{l23}$ to the ratio of the leptonic decays
\begin{equation}
R_{l23} = \frac{BR(K\to\mu\nu_{\mu})}{BR(\pi\to\mu\nu_{\mu})}\Big|_{MSSM}.
\end{equation}

 Recent work \cite{Bona:2009cj} has investigated the implications for the large $\tan\beta$ scenario of a new determination of the Standard Model prediction of the branching ratio $BR(B\to\tau\nu)$. The new value disfavours any supersymmetric contribution to this process, except when $\tan\beta$ is large and the charged Higgs mass $m_{H^+}$ is small. This situation is strongly constrained by $BR(B\to
X_s \gamma)$ and $BR(B_s\to\mu^+ \mu^-)$. Furthermore, the anomalous magnetic moment of the muon favours some supersymmetric contribution to achieve agreement with experiment. Thus, we expect some tension between $(g-2)_{\mu}$ and some of the $B$ observables.

In order to investigate this we now turn to a $\chi^2$ analysis. The $\chi^2$ value of the $i^{th}$ observable is
\begin{equation}
\chi^2_i = \frac{(p_i - c_i  )^2}{\sigma^2_i}
\label{eq:chi2}
\end{equation}
where $p_i$ is the predicted value and $c_i$ is the experimental central value.  This is not the case for the Higgs mass, for which we use a parametrisation of the LEP likelihood provided in the \texttt{SoftSUSY} package, and the unobserved branching ratio $BR(B_s \to \mu^+ \mu^-)$ where we use the Tevatron
likelihood\footnote{We thank C.~S. Lin for providing the likelihood for this process.}. The total $\chi^2_{tot} =\sum_i \chi_i^2$ is the sum of the $\chi^2$ values of the individual observables. We note that a study in similar spirit to ours has
been performed in the context of ordinary gauge mediation in~\cite{Gabrielli:1998sw, Gabrielli:1997jp}.

\begin{table}
\begin{center}
\begin{tabular}{|c|c|c|c|} \hline
Observable & Constraint & Experiment & Theory  \\ \hline \hline
$\delta a_{\mu}\times 10^{10}$ & $29.5\pm8.8$   & \cite{Amsler:2008zzb}    &\cite{Belanger:2008sj,Belanger:2006is,Belanger:2004yn,Belanger:2001fz, Heinemeyer:2004yq, Heinemeyer:2003dq,Marchetti:2008hw} \\ \hline

$m_h$[GeV]              & $>114.4$~GeV & \cite{Barate:2003sz}  &\cite{Allanach:2004rh} \\ \hline

$BR(B\to X_s \gamma)\times 10^4$ & $ 3.28\pm 0.29 $ &  \cite{Barberio:2008fa} &\cite{Mahmoudi:2007vz, Mahmoudi:2008tp}   \\
\hline
$BR(B_s\to\mu^+\mu^-)$ & $<5.8\times10^{-8}$ &  \cite{:2007kv} &\cite{Belanger:2008sj,Belanger:2006is,Belanger:2004yn,Belanger:2001fz} \\ \hline
$BR(B\to D \tau \nu)$   &$0.416\pm0.138$ & \cite{Aubert:2007dsa}  &\cite{Mahmoudi:2007vz, Mahmoudi:2008tp} \\ \hline
$BR(D_s \to \tau\nu)$ & $5.7\pm0.5 \times 10^{-2}$ & \cite{Akeroyd:2009tn}  &\cite{Mahmoudi:2007vz,Mahmoudi:2008tp} \\ \hline
$BR(D_s \to \mu\nu)$ & $5.7\pm0.5\times10^{-3}$ & \cite{Akeroyd:2009tn} & \cite{Mahmoudi:2007vz,Mahmoudi:2008tp} \\ \hline
$R_{B\tau\nu}$      & $1.9\pm0.60$ &   \cite{Bona:2009cj} &\cite{Mahmoudi:2007vz,Mahmoudi:2008tp,Isidori:2006pk} \\ \hline
$\Delta_{0-}$      & $0.031^{+0.03}_{-0.025}$ & \cite{Amsler:2008zzb,:2008cy, Nakao:2004th}& \cite{Mahmoudi:2007vz, Mahmoudi:2008tp}  \\ \hline
$R_{l23}$               &$1.004\pm0.007$ & \cite{Antonelli:2008jg}  & \cite{Mahmoudi:2007vz,Mahmoudi:2008tp} \\ \hline \hline

\end{tabular}
\end{center}
\caption{Experimental constraints, showing the observables, the constraints applied and the source of the theoretical and experimental values and errors.}
\label{tab:obs}
\end{table}

\begin{figure}
\vspace*{-0.6cm}
\begin{center}
\subfigure[]{
\includegraphics[bb= 142 78 510 410,clip,width=6.5cm]{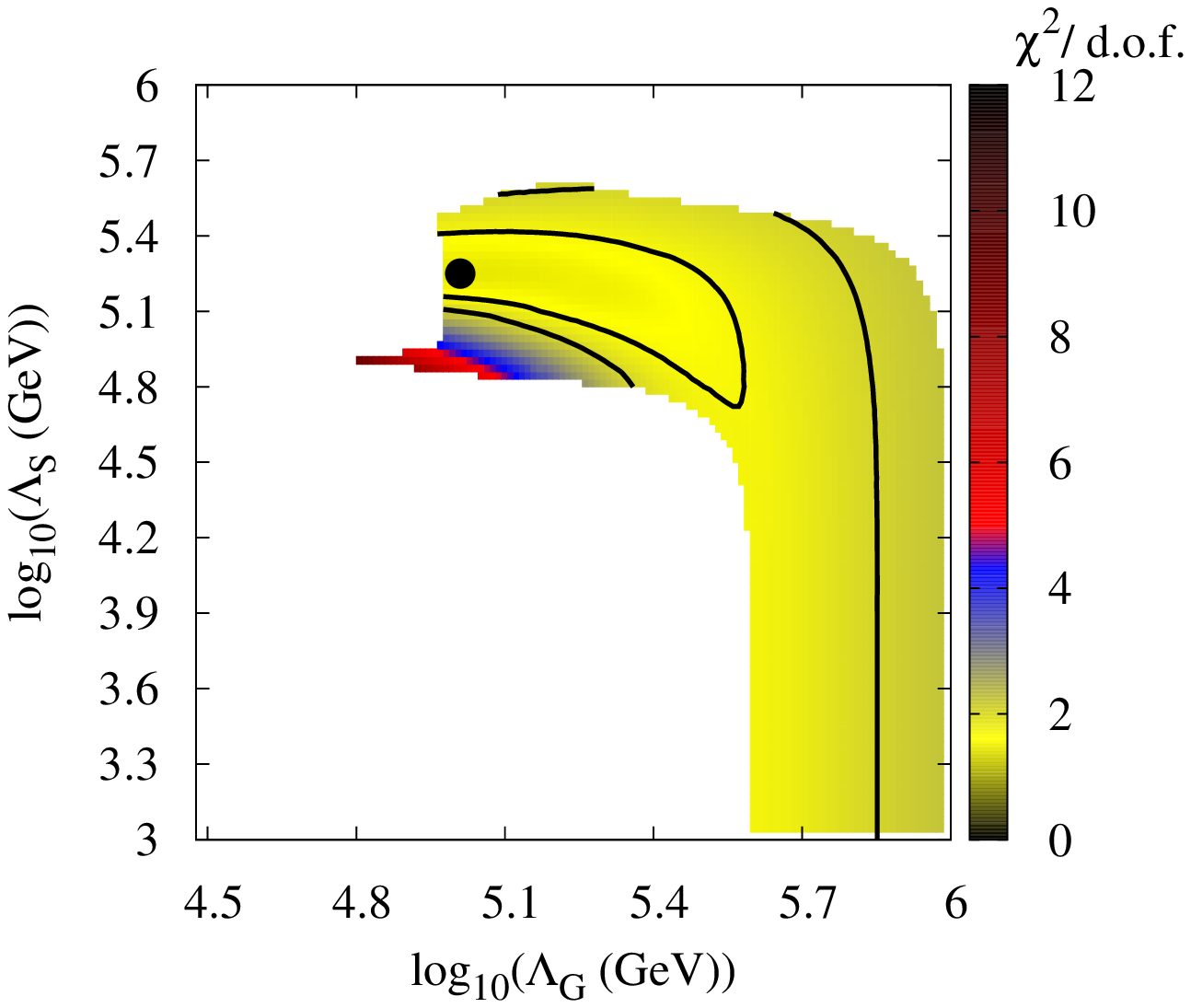}
}
\hspace*{1cm}
\subfigure[]{
\includegraphics[bb= 142 78 510 410,clip,width=6.5cm]{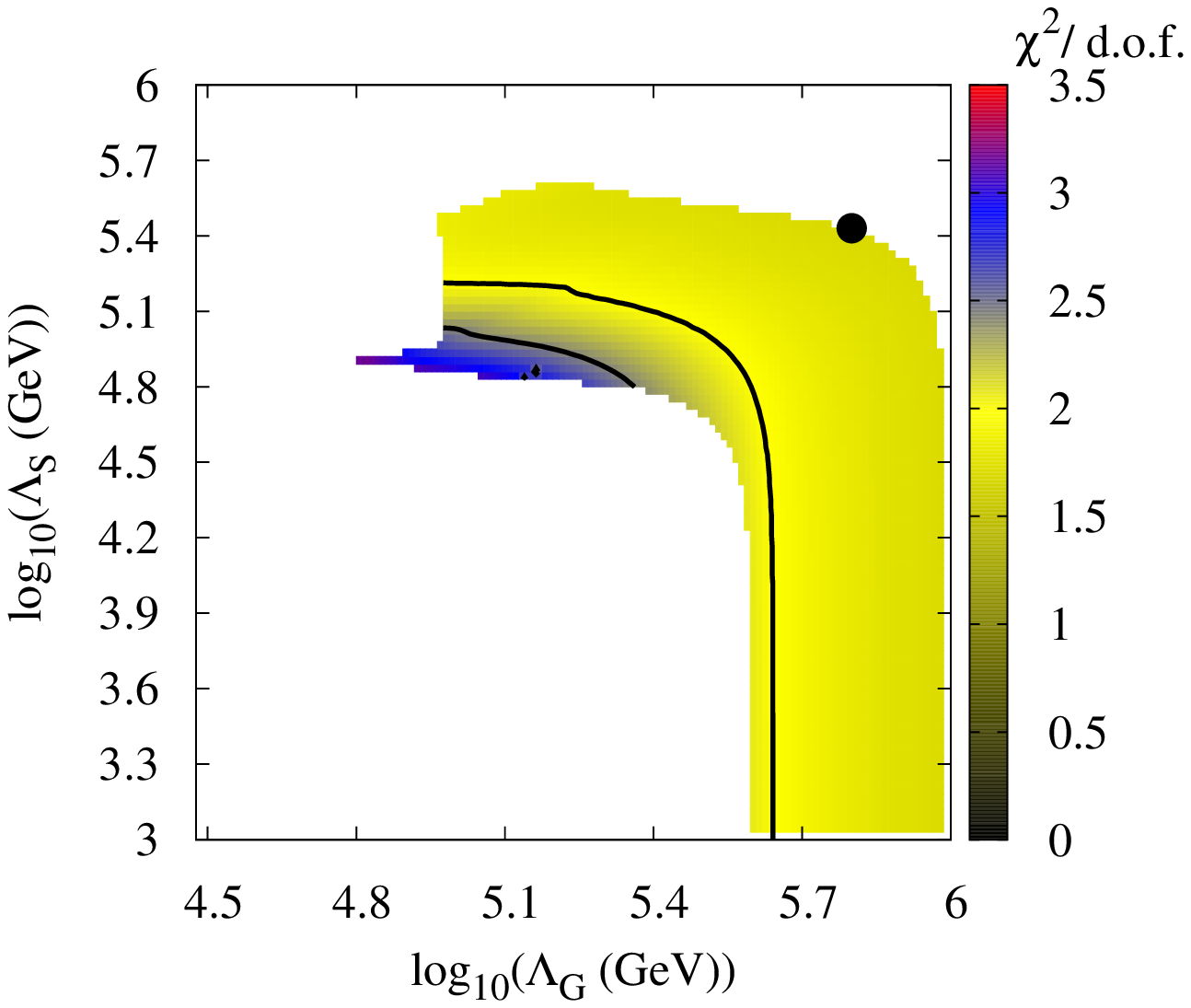}
}
\\
\vspace*{-0.7cm}
\subfigure[]{
\includegraphics[bb= 142 78 510 410,clip,width=6.5cm]{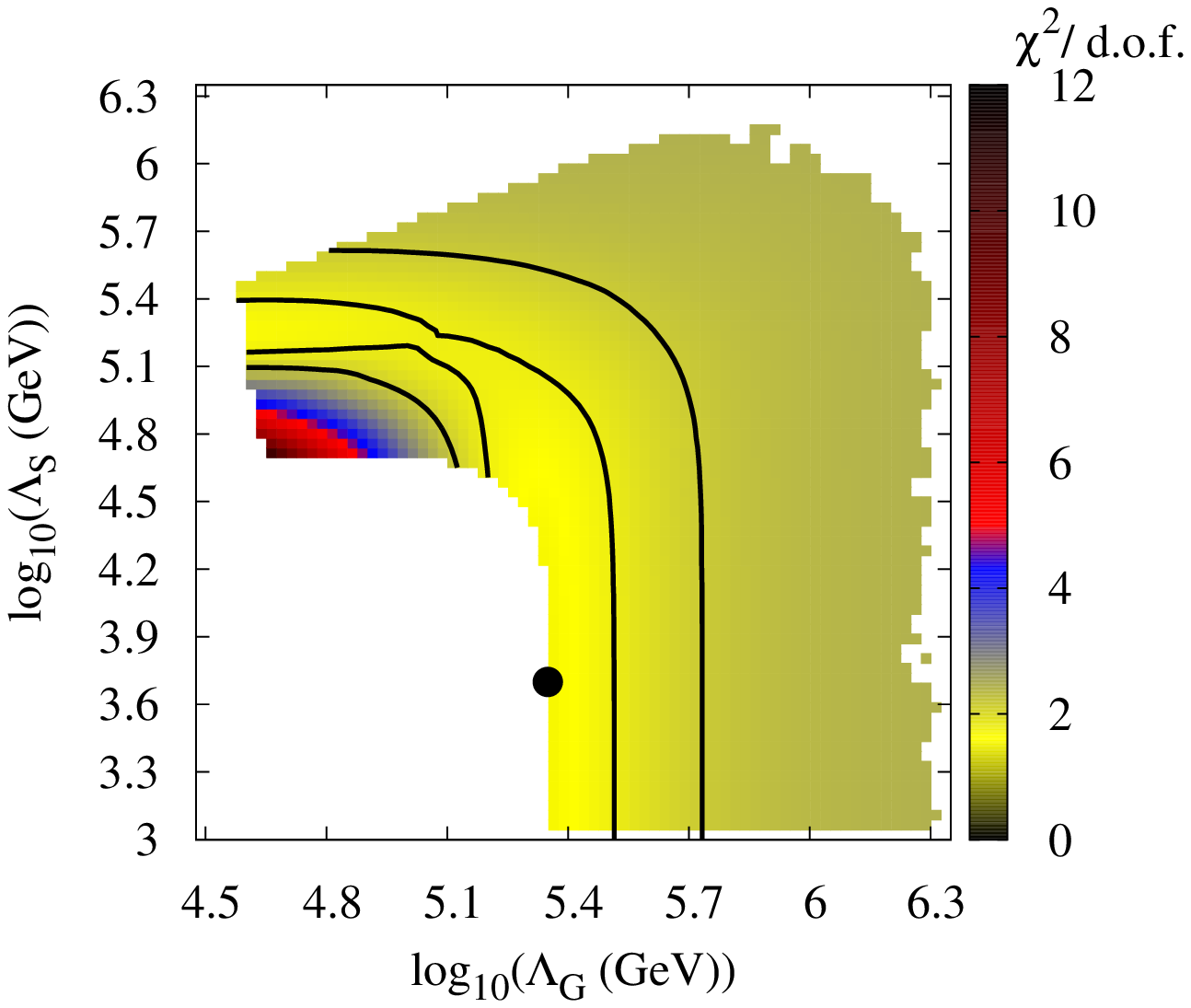}
}
\hspace*{1cm}
\subfigure[]{
\includegraphics[bb= 142 78 510 410,clip,width=6.5cm]{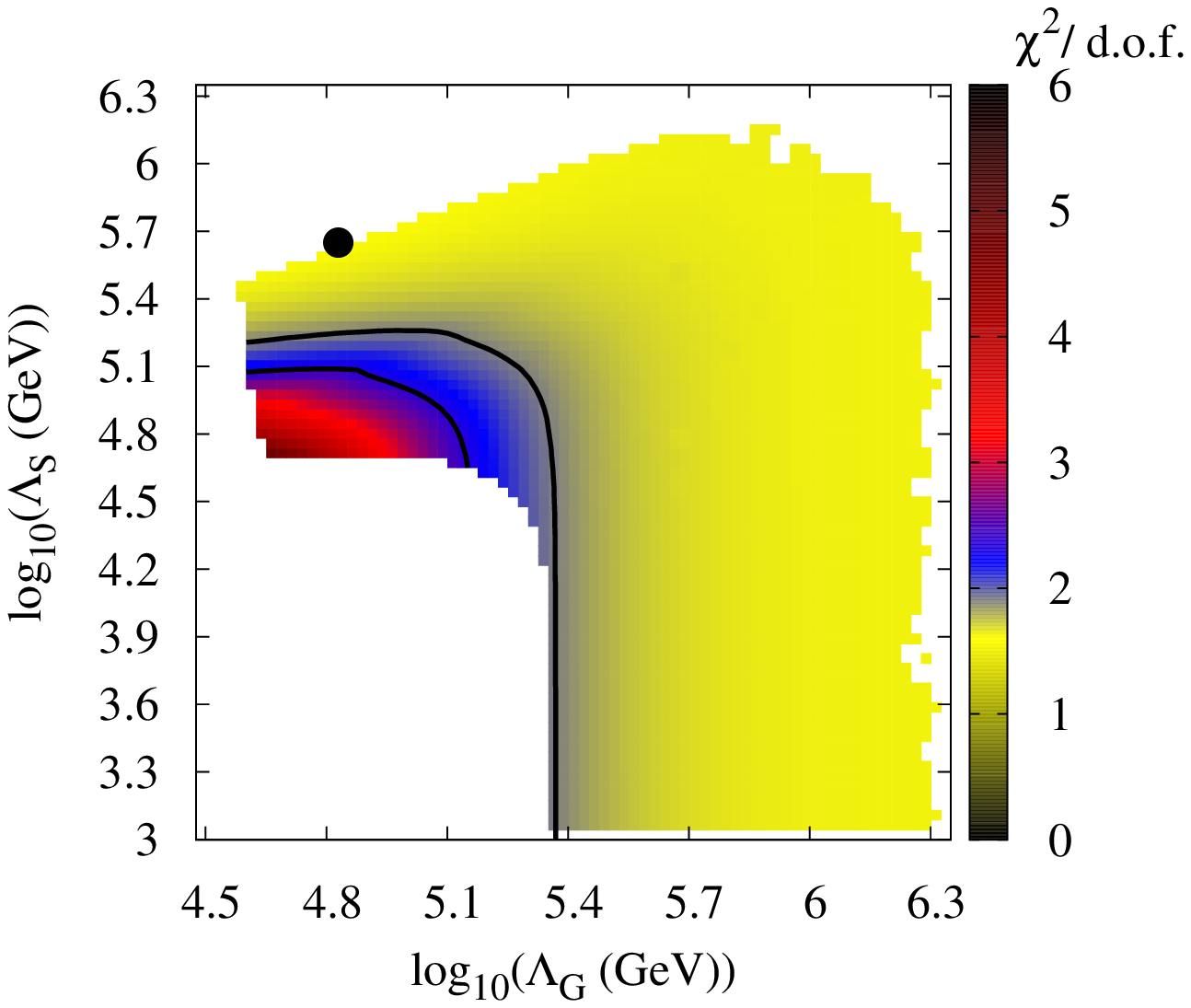}
}
\\
\vspace*{-0.7cm}
\subfigure[]{
\includegraphics[bb= 142 78 510 410,clip,width=6.5cm]{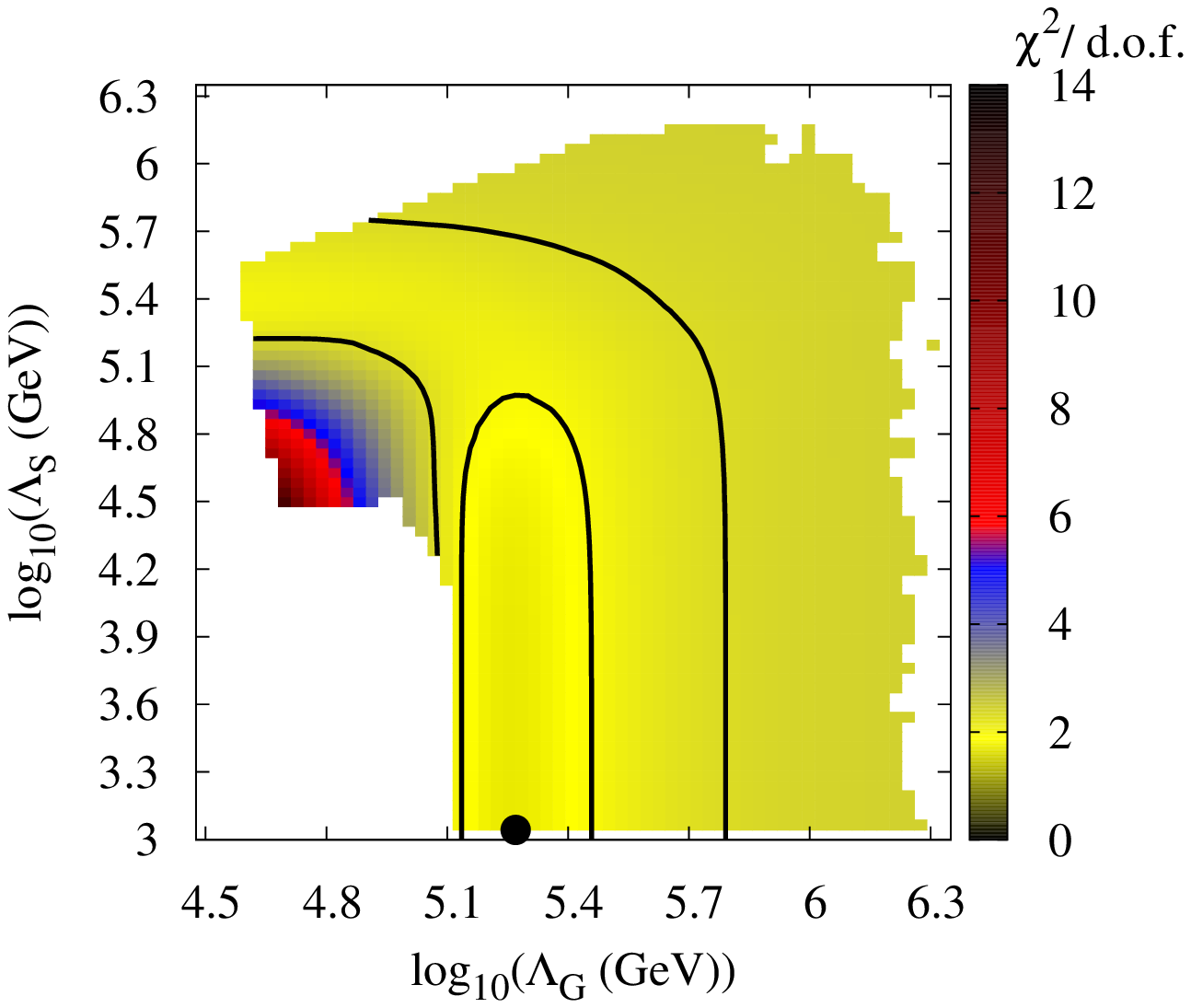}
}
\hspace*{1cm}
\subfigure[]{
\includegraphics[bb= 142 78 510 410,clip,width=6.5cm]{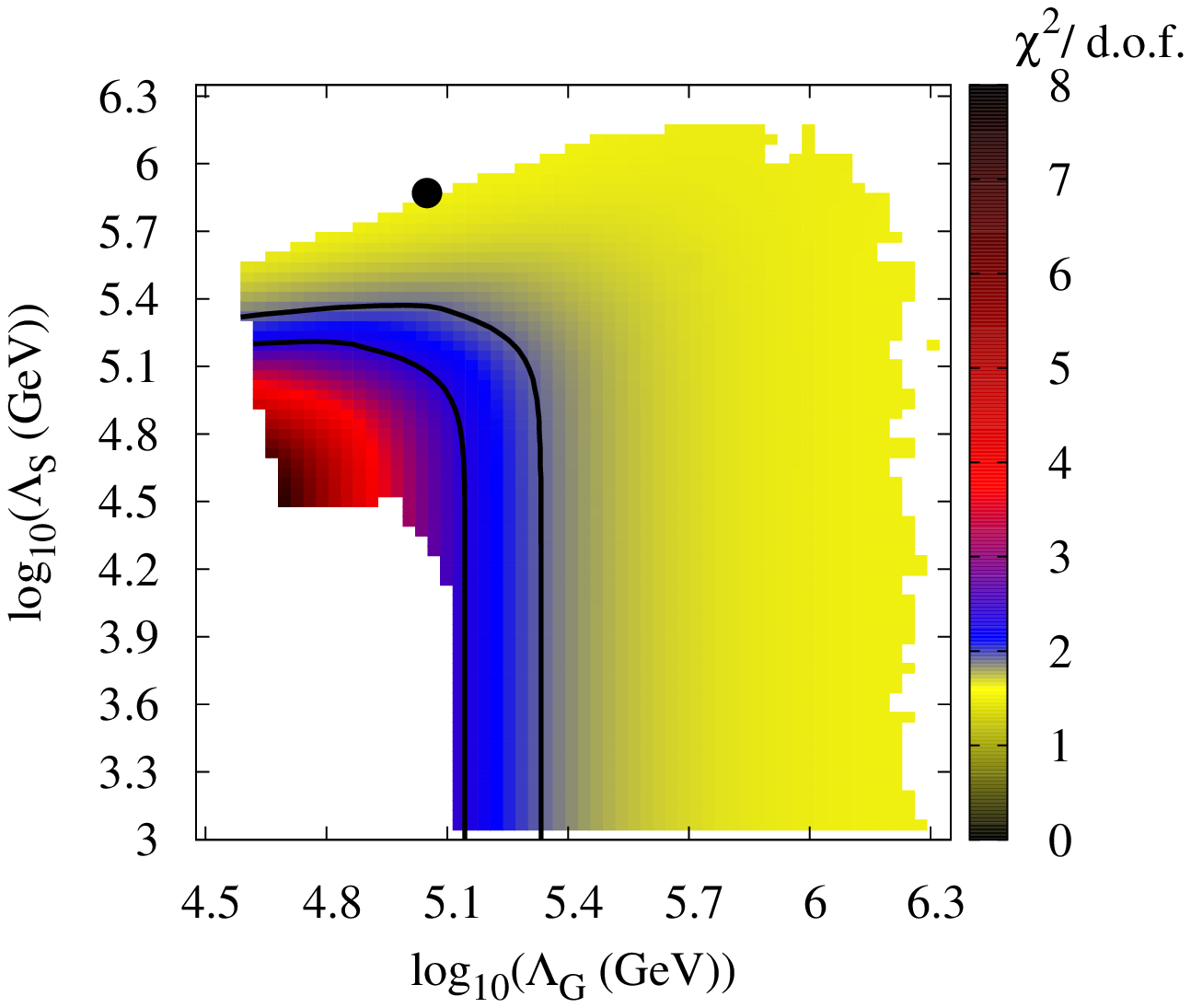}
}
\caption{
 (a,c,e) show the $\chi^2_{tot}$ distribution in the $\Lambda_G$-$\Lambda_S$ plane for $M_{mess}=10^6$ ,$10^{10}$
and $10^{14}$~GeV respectively, and (b,d,f) show the $\chi^2$ of only the B physics
observables for the same values of $M_{mess}$. The black lines denote the boundaries of the $68\%$ and $95\%$ confidence regions.
The black spots mark the best-fit points in all cases.}
\label{fig:fits}
\end{center}
\end{figure}

Figure \ref{fig:fits} (a,c,e) show the $\chi^2_{tot}/d.o.f.$ distributions
we obtain from the scans for $M_{mess}=1\times 10^{6,10,14}$ respectively, along with 68\% and 95\% confidence
limit contours ($\Delta\chi^2_{tot} = 2.41, 5.99$ respectively).
The region of maximum likelihood is shown in yellow, and the best-fit points are marked by black splodges. We see immediately that the region of light supersymmetry where both $\Lambda_G$ and $\Lambda_S$ are small is strongly disfavoured (the blue and red region). This due to a combination of factors. Since the scalars are light the Higgs mass is below the LEP bound for which there is a strong $\chi^2$ penalty. On top of that, the supersymmetric contributions to $(g-2)_{\mu}$ and the $B$-observables are too large. As the masses of the SUSY particles increase the loop contributions become smaller and the Higgs mass larger.
A large amount of the region of good fit for the higher $M_{mess}$
has very small $\Lambda_S$ for $M_{mess}=10^{10}$~GeV, and all of the $68\%$ confidence region for $M_{mess}=1\times 10^{14}$ has an inverted hierarchy $\Lambda_G > \Lambda_S$.
For all values of $M_{mess}$ within the 68\% confidence limits shown the Higgs mass is just above the limit set at LEP, and the anomalous magnetic moment of the muon is saturated by SUSY effects. However, the region of best-fit to the combined observables is not the best-fit region of the combined $B$-observables. We have found the most sensitive of the $B$-observables to be $BR(B\to X_s \gamma)$ and $BR(B\to\tau\nu)$. As discussed in \cite{AbdusSalam:2009tr,Trotta:2008bp} there is a tension between the SUSY contributions to $(g-2)_{\mu}$ and $BR(B\to X_s \gamma)$. The current Standard Model prediction of $BR(B\to X_s \gamma)$ is consistent with the experimental value. Also, while there is a small discrepancy between the SM prediction and the experimental determination of $BR(B\to\tau\nu)$, any supersymmetric contribution to this process will lead to a worse fit than the Standard Model\footnote{This is not necessarily true if $\tan\beta$ is very large and $m_{H^+}$ is very small. In our scenario $m_{H^+}$ is never small enough for this to happen.}. These data thus pull towards the decoupling limit of a heavy supersymmetric spectrum. We can see this in Figure \ref{fig:fits} (b,d,f)
which show the $\chi^2$ distribution and 68\% and 95\% confidence limits obtained when we omit $m_h$ and $(g-2)$ from the total $\chi^2$
for $M_{mess} = 10^{6,10,14}$ respectively. Dark spots mark the new best-fit points. There is a clear preference for large values of $\Lambda_G$ and $\Lambda_S$. While the mass of the lightest Higgs is well above the lower bound and thus contributes nothing to the $\chi^2_{tot}$, in this region the SUSY contributions to $(g-2)_{\mu}$ are very small, leading to this observable having an individual $\chi^2$ between 5 and 11 depending on the details of the spectrum.

Considering again the case of all observables combined, the $\chi^2_{tot}$ value of
the best-fit points are 14.55, 15.47, 17.08 and for $M_{mess}= 10^6$, $10^{10}$ and $10^{14}$~GeV respectively. These correspond to $p$-values of 0.069,0.051 and 0.029. If we adopt a significance level of 5\%, then we would reject the possibility that $M_{mess}=1\times 10^{14}$~GeV.
 The values of $\Lambda_G$, $\Lambda_S$ and the constraints at the best-fit points are shown in Table \ref{tab:bestfit}. We provide the spectra of these three points in Appendix
\ref{sec:Spectra}.
\begin{table}
\begin{center}
\begin{tabular}{|c||c|c||c|c||c|c|} \hline
$M_{mess}$ & \multicolumn{2}{|c||}{$10^{6}$~GeV} & \multicolumn{2}{|c||}{$10^{10}$~GeV} & \multicolumn{2}{|c|}{$10^{14}$~GeV} \\
\hline
\hline
$\Lambda_G$ (GeV) & \multicolumn{2}{|c||}{$1.08\times 10^5$} & \multicolumn{2}{|c||}{$2.25\times 10^5$} &
\multicolumn{2}{|c|}{$1.86\times 10^5$}  \\ \hline
$\Lambda_S$ (GeV) & \multicolumn{2}{|c||}{$1.79\times 10^5$} & \multicolumn{2}{|c||}{$4.96\times 10^3$}
 & \multicolumn{2}{|c|}{$1.11\times 10^3$}     \\ \hline
$\tan\beta$      & \multicolumn{2}{|c||}{57.5} & \multicolumn{2}{|c||}{25.2} & \multicolumn{2}{|c|}{21.5} \\ \hline
$\chi^2_{total}$  & \multicolumn{2}{|c||}{14.55}    & \multicolumn{2}{|c||}{15.47}   & \multicolumn{2}{|c|}{17.08} \\ \hline
 & Value & $\chi_i^2$ & Value & $\chi_i^2$ & Value & $\chi_i^2$ \\ \hline
$\delta a_{\mu}=(g-2)_{\mu}|_{\rm MSSM}\times 10^{-10}$ & 26.8 & 0.1 & 20.6 & 1.0  & 17.8 & 1.8 \\ \hline
$m_h$ (GeV)               & 118 & -1.0 & 117 & -1.0 & 117 & -1.0  \\ \hline
 $BR(B\to X_s\gamma)\times 10^{-4}$ & 3.26 & 1.1 & 3.05 & 3.5 & 3.00 & 4.0\\ \hline
 $R_{B\tau\nu}$ & 0.63 & 4.5 & 0.95 & 2.5 & 0.96 & 2.4 \\ \hline
 $\Delta_{0-}$ & 0.078 & 2.5 & 0.082 & 2.9  & 0.083 & 3.0\\ \hline
 $BR(B_s \to \mu^+\mu^-)\times 10^{-9}$ & 4.7 & 0 & 3.5 & 0 & 3.4 & 0 \\ \hline
 $BR(B\to D\tau\nu)$ & 0.27 & 1.2 & 0.29 & 0.8 & 0.30& 0.8 \\ \hline
\end{tabular}
\end{center}
\caption{Values of $\Lambda_G$, $\Lambda_S$ and the constraints at the best-fit points for $M_{mess}=10^6$~GeV (left column),
$M_{mess}=10^{10}$~GeV (middle column) and $M_{mess}=10^{14}$~GeV.}
\label{tab:bestfit}
\end{table}
More importantly for us, what are the implications for efforts towards model building? If we take the 95\% confidence limits as some indication of what the scales of the gaugino and scalar masses
should be, for all values of $M_{mess}$ we have
presented there is an upper bound on $\Lambda_S$ of around 500~TeV. What is perhaps more interesting is that we have not found it possible to put a lower bound on $\Lambda_S$.
Looking at Figure \ref{fig:fits} it is clear that $\Lambda_G / \Lambda_S \ll 10^{-1}$
is disfavoured by current indirect observables.
However, a mild hierarchy $\Lambda_G/\Lambda_S\gtrsim 1/10$ is consistent with the data.
This puts strong constraints on
models of pure direct gauge mediation\footnote{Strictly speaking direct mediation
models often have additional light messenger fields. In our numerical analysis
we have neglected their possible contributions to the running.}.
However, this ratio is less of a problem for hybrid models and models with explicit messengers.
What is more
surprising
is that $\Lambda_G/\Lambda_S$ can be as large as a few hundred
without being disfavoured for large values of the messenger scale.

It must also be stated that our fits apply only to the specific solution of the $\mu/B_{\mu}$ problem considered in this paper, which leads to large $\tan\beta$.  A model with large $\tan\beta$ can successfully fit $(g-2)_{\mu}$ with a heavier sparticle spectrum than one with smaller $\tan\beta$. On the other hand, smaller values of $\tan\beta$ are more
favoured by the B physics constraints.

\section{Fine Tuning and Naturalness Bounds}
\label{sec:finetuning}

The mass of the Z boson is obtained in the MSSM by minimization of the scalar Higgs
potential. At tree level this leads to the relationship
\begin{equation}
\frac{M_Z^2}{2} = \frac{m^2_{H_d} - m^2_{H_u} \tan^2\beta}{\tan^2\beta -1} - \mu^2
\label{eq:ewsb}
\end{equation}
 Eq.~(\ref{eq:ewsb}) is still valid under the full effective potential $V_{eff} = V_{tree} + \Delta V$ if one makes the substitutions
\begin{equation}
m_{H_{u,d}}^2  \to m_{H_{u,d}}^2 + \frac{1}{2v_{u,d}} \frac{\partial (\Delta V)}{\partial v_{u,d}}
\end{equation}
where the vacuum expection values $v_{u,d}$ are treated as real parameters in the differentiation.
In the relevant limit of large $\tan\beta$ this becomes
\begin{equation}
\frac{M_Z^2}{2} = -\left( m_{H_u}^2 + \mu^2 \right) + \frac{1}{\tan^2\beta} \left( m_{H_d}^2 -m_{H_u}^2 \right)
+ \mathcal{O}(1/\tan^4\beta)
\label{eq:ewsb_tanb_exp}
\end{equation}

Therefore if large cancellations do not take place, $-m_{H_u}^2$ and $\mu^2$ should be of the same order of magnitude as $M_Z^2$.

At tree-level the mass of the lightest Higgs is below $M_{Z}$.
It receives large one-loop corrections from the top sector, which allow
for phenomenologically acceptable values of the Higgs mass if the stops are
sufficiently heavy.
A recent detailed study, Ref.~\cite{Essig:2007vq},
gives $M_S=\sqrt{m_{\tilde{t}_{1}}m_{\tilde{t}_{2}}} \geq 600$~GeV.
Further constraints on the spectrum come from the direct search limits at LEP and Tevatron.
The large lower bounds obtained from these experiments
on the scalar masses implies that some cancellations must
take place in Eq.~(\ref{eq:ewsb}) for the Z boson to have the correct mass.
Such cancellations, where it is necessary to fine-tune one of the parameters of
the theory in order to achieve a phenomenologically acceptable result,
are not desirable in a theory specifically proposed for its ability to
solve one of the major fine-tuning problems of the Standard Model.
As direct search limits increase, the minimum allowed fine-tuning increases,
decreasing our belief that supersymmetry is realised. Furthermore there is some
inherent subjectivity in how one chooses to define an appropriate measure of
fine-tuning, and what constitutes an acceptably high level of fine-tuning in a theory. Should we accept fine-tuning at the $10\%$ level, but not $1\%$? For these reasons we think that while an analysis of the necessary fine-tuning required to achieve electroweak symmetry breaking in General Gauge Mediation is worthwhile, arguments based on fine-tuning should not be used to rule out any theory under consideration.

\begin{figure}
\vspace*{-0.6cm}
\begin{center}
\subfigure[]{
\includegraphics[bb= 142 78 510 410,clip,width=6.5cm]{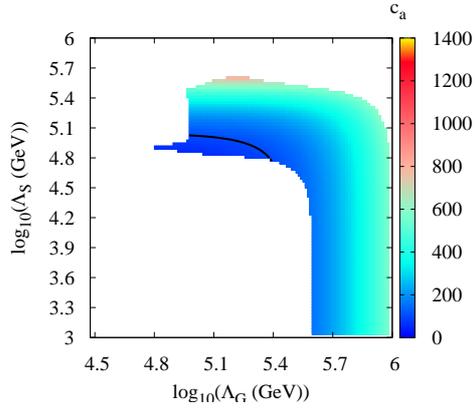}
}
\end{center}
\vspace*{-1.2cm}
\begin{center}
\subfigure[]{
\includegraphics[bb= 142 78 510 410,clip,width=6.5cm]{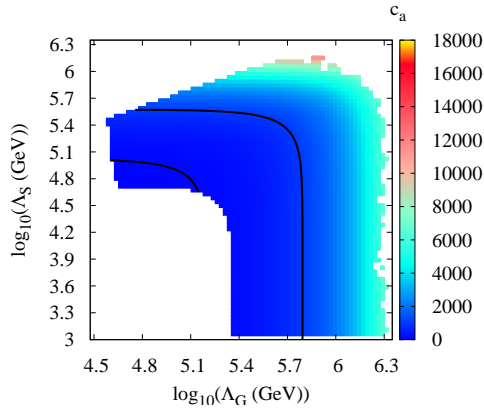}
}
\end{center}
\vspace*{-1.2cm}
\begin{center}
\subfigure[]{
\includegraphics[bb= 142 78 510 410,clip,width=6.5cm]{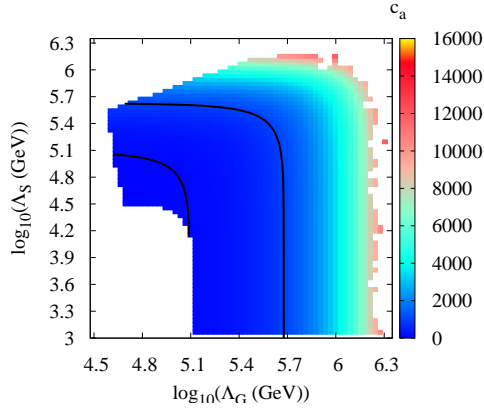}
}
\end{center}
\begin{center}
\vspace*{-0.3cm}
\caption{Plots showing level of fine-tuning required to successfully break electroweak symmetry, $c_a$ (roughly speaking we have to tune to 1 part in $c_{a}$) for
(a) $M_{Mess} = 10^{6}$~GeV, (b) $M_{Mess} = 10^{10}$~GeV and (c) $M_{mess}= 10^{14}$~GeV. Also shown are the contours of
$c_a =100$ and $1000$.}
\label{fig:FT}
\end{center}
\end{figure}

 With this caveat in place let us proceed. A number of definitions of a suitable quantification of fine-tuning have been proposed
\cite{Barbieri:1987fn,deCarlos:1993yy}. In this work we adopt the definition of \cite{deCarlos:1993yy},
which is incorporated in the \texttt{SoftSUSY} code. Consider a set of model parameters $\{a\}$. For us $a = \{\Lambda_G, \Lambda_S, \mu\}$.
Since $B_{\mu}$ is set to be zero at the high scale $M_{mess}$
we do not, of course, consider it to be part of our fine-tuning measure. Then the sensitivity of $M_Z^2$ to the parameter $a_i$ is
\begin{equation}
c_{a_i} \equiv \left| \frac{\partial \ln M_Z^2}{\partial \ln a} \right|
\end{equation}
The total fine-tuning in the soft-parameters is defined to be $c_a = \max(c_{a_i})$.
While this makes clear the sensitivity of $M_Z^2$ to the soft breaking parameters, it is also possible that there could exist a
region of parameter space
that
evades the naturalness bounds in the soft-parameters but is very finely tuned with respect to some other parameter. The canonical example of this is the focus-point region in
the Constrained MSSM~\cite{Romanino:1999ut}, which despite being a region of low
fine-tuning from the perspective of the universal scalar mass $m_0$, is nonetheless
very sensitive to the top Yukawa $h_t$. The top Yukawa coupling is different
in some ways to the soft masses we have included in our definition of $c_a$: it
is dimensionless and is related in an intimate way to the measured mass of the top
quark $M_t$~\cite{Ross:1992tz}. For these reasons we do not include it in
our definition of $c_a$.
We have examined the results for $c_{h_t}$ and found them to be
qualitatively similar.

We show in Figure \ref{fig:FT} (a,b,c) the level fine-tuning required
in our scans with $M_{mess} = 10^{6}$, $10^{10}$ and $10^{14}$~GeV as in the previous section
\footnote{For earlier work on alleviating the fine-tuning problem in gauge mediation see \cite{Agashe:1997kn}}.
We also show contour
lines corresponding to fine-tuning at the $c_a = 100$ and
$1,000$ levels. The minimum fine-tuning possible is around $c_a \sim 30$.
However, the region in which this occurs is strongly
disfavoured by the low energy observables.
In fact, the region
preferred by the low energy observables is quite well delineated by
the contours of $100$ and $1,000$ for $M_{mess}=1\times 10^{10,14}$. That this amount of fine-tuning is
necessary is somewhat
troubling, but is comparable to the situation in the
mSUGRA scenario.

It is interesting to note that the $\chi^2$ (Figure~\ref{fig:fits})
and the fine-tuning (Figure~\ref{fig:FT}) prefer different regions of parameter space.
Without the $\chi^2$ analysis the fine-tuning plots alone would favour light
supersymmetry.
However, the comparison with measured observables, i.e. the $\chi^2$,
favours somewhat heavier superpartner masses.

Finally let us
comment on the $\Lambda_G/\Lambda_S$ ratio and its effect on the
amount of fine-tuning.
Keeping $\Lambda_S$ fixed and moving horizontally to the left, i.e. decreasing $\Lambda_G$
we see that the fine-tuning decreases.
This is because the fine-tuning is dominated by the scalar mass squareds which decrease
when either $\Lambda_G$ or $\Lambda_S$ decrease.
In particular, beginning on the line of ordinary gauge mediation
and decreasing $\Lambda_G$ (with low $\chi^2$) does not lead to a significant increase
in either the amount of fine-tuning or in $\chi^2$.
In this sense mildly split SUSY is not at all disfavoured compared
to ordinary gauge mediation\footnote{Conversely, if we were to reduce the $\Lambda_G/\Lambda_S$ ratio
by increasing $\Lambda_S$ with $\Lambda_G$ fixed we \emph{would} see
an increase in the necessary fine-tuning. The second procedure is essentially an
increase in the supersymmetry breaking scale whereas the first procedure can be
viewed as keeping the SUSY-breaking scale fixed but increasing the
amount of direct mediation.}.

\section{Conclusion}
We have investigated the phenomenology of pure general gauge mediation.
In this setup the $B_{\mu}$ parameter is close to zero at the messenger scale
but appears at low energies due to running.
Consequently its value is rather small at the electroweak scale, leading to relatively
large values of $\tan\beta\sim 15-65$.
In the spirit of general gauge mediation we treat the scalar and gaugino mass
scales, $\Lambda_S$ and $\Lambda_{G}$, as two \emph{independent} input parameters.
We applied a raft of experimental constraints including the Higgs mass, $(g-2)_{\mu}$
and $B\to X_s\gamma$. We determined the favoured region of parameter space
which includes mildly-split ($\Lambda_G\sim 0.1 \Lambda_S$) as well as
the non-split ($\Lambda_G=\Lambda_S$) SUSY signatures, characteristic of direct/hybrid and
ordinary gauge mediation, respectively. Our $\chi^2$ analysis does not favour
one over the other within the phenomenologically preferred region.
The opposite hierarchy $\Lambda_G\gg\Lambda_S$ is also allowed.
We find lower messenger masses to be slightly favoured over higher ones.
The fine-tuning is typically of the percent level.

General gauge mediation and gravity mediation both have rich parameter spaces.
Therefore it is useful for phenomenological analyses to define a canonical model
that
encompasses the most relevant features. In the case of gravity mediation
this has always been mSUGRA/CMSSM with the parameters $m_{0}$ and $m_{1/2}$
(as well as $\tan\beta$ and $A_{0}$).
Here we have analysed a similar characterisation
of general gauge mediation, pure GGM, in terms of $\Lambda_S$, $\Lambda_G$ and $M_{mess}$
(the $B_{\mu}$ parameter and trilinear couplings being determined to be small at the messenger
scale).

\section*{Acknowledgements}
We are grateful to Ben Allanach, Callum Durnford and Frank Krauss for useful discussions.
SAA and VVK are
in receipt of Leverhulme Research Fellowships. MJD thanks St John's College
Cambridge and EPSRC for financial support.
We thank the CERN theory group and the IPPP for hospitality.

\startappendix

\section{Generating $\mu$ and $B_{\mu}$ in GGM}\label{app:bmu}

The authors of~\cite{Komargodski:2008ax} discussed two types of (non-gauge)
couplings between the SUSY-breaking sector and the Higgs fields.
The first type involves $SU(2)$ doublet hidden-sector operators $\Phi_{u,d}$
which couple to the Higgs fields through the superpotential coupling
 \be
 \label{doubcoup}
 \int d^2 \theta \, \left( \lambda_u \CH_u \Phi_d + \lambda_d \CH_d \Phi_u\right) \ ,
 \ee
and the second class of models is in terms of an $SU(2)$ singlet
hidden-sector superfield $\CS$, which couples to the Higgs fields as
 \be
 \label{singcoup}
 \int d^2 \theta\,  \lambda^2 \CS \CH_u \CH_d \ .
 \ee
The hidden-sector fields $\Phi_{u,d}$ in \eqref{doubcoup} and $\CS$
in \eqref{singcoup} can be composite operators of an
underlying theory or could be elementary fields.
We shall always assume that these models are
weakly coupled, i.e.  $\lambda_u, \lambda_d, \lambda \ll 1$.

In the following we will analyse the doublet model \eqref{doubcoup}.
The singlet model can be analysed in a very similar way and the results are
qualitatively similar.

These models are further divided into two categories depending on how
many distinct scales exist in their hidden sectors.

\subsection{One-scale models}

This is the case (c) from the main text where the hidden-sector is characterized by a
single effective scale $F/M$,
where $F$ is the SUSY-breaking $F$-term vev and $M$ is the messenger mass.

To leading order in the couplings $\lambda_u$ and $\lambda_d$ and in the SUSY-breaking
scale,
the model with doublet messengers in \eqref{doubcoup} generates the following scaling pattern
of contributions to $\mu$ and the soft terms~\cite{Komargodski:2008ax}
\bea
\label{effonescale}
& \mu\sim \lambda_u\lambda_d {F\over M}\\
& B_\mu\sim \lambda_u\lambda_d {F^2\over M^2} \nonumber\\
& \delta m^2_{{u,d}}\sim \left|\lambda_{u,d}\right|^2{F^2\over M^2}\nonumber\\
& \delta a_{u,d}\sim \left|\lambda_{u,d}\right|^2{F\over M}\nonumber\,.
\eea
Since both $\mu$ and $B_\mu$ are generated at order $\lambda_u\lambda_d$
and there is only one mass-scale in the problem, it follows that
$B_\mu /\mu^2 \sim 1/(\lambda_u\lambda_d) \gg 1$, thus giving the standard incarnation of the $\mu/B_\mu$-problem
which precludes electroweak symmetry breaking.
To improve the situation, one would have to suppress the $B_\mu$ term
by higher powers of small hidden-sector couplings, which can be achieved, as
discussed in
Refs.~\cite{Dvali:1996cu,Dine:1996xk,Yanagida:1997yf,Dimopoulos:1997je,Langacker:1999hs,Hall:2002up,Giudice:2007ca,Liu:2008pa},
but requires a specially engineered SUSY-breaking sector.

\subsection{Models with more than one scale}

Now the hidden-sector is characterized by at least two mass scales.
Essentially we assume that some fields of the hidden-sector acquire vevs
at the scale $M$,
but supersymmetry is broken at a much lower energy with
$F \ll M^2$. We treat $F$ and $M$ as independent parameters, so that $M$ does not vanish
in the limit of restored SUSY $F \to 0$.
Since $\mu$ is a supersymmetric parameter, it does not depend on $F$ to leading order
in two-scale models, but the SUSY-breaking terms, such as $B_\mu$ must depend on $F$.
More generally, these models can accommodate a variety of messenger scales (different messenger masses
is a typical feature of dynamical SUSY-breaking models with direct gauge mediation).
In particular we can have a structure $\sqrt{F} < M_{mess} < M$, where $M_{mess}$ is a typical mass
of messengers participating in the pure gauge mediation, and $M$ is a vev scale of doublet messengers
interacting with Higgses.

To leading order in $\lambda_u$ and $\lambda_d$ and in the SUSY-breaking
scale $F$,
the model which included doublet messengers \eqref{doubcoup} can express
a typical scaling pattern \cite{Komargodski:2008ax}
\bea
\label{twscale}
&&\mu\sim \lambda_u\lambda_d M   \\
&& B_\mu\sim c^{(1)}\lambda_u\lambda_d F \,+ c^{(2)}\lambda_u\lambda_d \, {F^2\over M^2_{mess}} \nonumber\\
&&\delta m^2_{{u,d}}\sim \left|\lambda_{u,d}\right|^2{F^2\over M^2_{mess}}\nonumber\\
&& \delta a_{u,d}\sim \left|\lambda_{u,d}\right|^2{F\over M_{mess}}\nonumber\\
&& {F \over M^2_{mess}} \ll 1 \ , \quad \lambda_u \sim \lambda_d \ll 1
\nonumber
\eea
In the equation for $B_\mu$ we have shown the leading and the subleading term in the SUSY-breaking scale $F$.
The authors of \cite{Komargodski:2008ax} were strongly attracted to using symmetry
to suppress the leading order contribution to $B_\mu$ thus setting $c_1 =0$ and $c_2 \sim 1$.
This is indeed possible in the class of models they considered, and
the symmetry reasoning can be based
on the fact that the effective $\mu$ and $B_\mu$ terms in \eqref{mudef}, \eqref{quaddef} cannot be both neutral
under the $R$-symmetry (and the same applies to Peccei-Quinn symmetry which rotates both $H$ fields by the same phase).

With this constraint in place one is able to consider simple two scale models with $M_{mess}\sim M$,
\be
\label{sublead}
B_\mu\sim \lambda_u\lambda_d \, {F^2\over M^2}\, .
\ee
If all the small parameters
of the system happen to be of the same order, i.e.
\be
\label{same}
{F\over M^2}\, \sim\,  \lambda_{u,d}\, \sim\,  \alpha_{\sst SM} \sim 10^{-3}-10^{-2} \,
\ee
one finds
\be
\mu^2 \sim B_\mu \sim \delta m^2_{{u,d}} \quad {\rm and} \quad \mu \gg \delta a_{u,d} \ ,
\ee
which implies for the $\lambda$-extension of
GGM:
\begin{enumerate}
\item{} both $\mu$ and $B_\mu$ are generated as inputs above the electroweak scale (so that $\mu^2 \sim B_\mu $
and there is no $\mu / B_\mu$ problem affecting electroweak symmetry breaking);
\item{} there are non-trivial contributions to the soft Higgs masses $\delta m^2_{{u,d}}$ on top of the
expected GGM contributions $\sim \alpha_{SM}^2$;
\item{} the additional (to GGM) contributions to the trilinear soft terms are negligible, $\delta a_{u,d} \simeq 0$.
\end{enumerate}
In particular point (2) marks a deviation from the spirit of gauge mediation.

We will now see what happens when we do not impose Eq.~\eqref{same}.
We can unfreeze the small parameters in \eqref{same} taking instead
\be
\label{notsame}
{F\over M^2}\, \ll \,  \lambda_{u,d}\, \ll \,  1 .
\ee
This can be achieved by increasing $M$ while keeping $F$ fixed. This really corresponds to
the case of $\mu$ being generated at a scale which has nothing to do with the
SUSY-breaking $F$.
Then
\be
\label{c2}
{B_\mu \over \mu^2} \, \sim \, {1 \over \lambda_u\lambda_d}\, {F^2 \over M^4} \, \ll 1 \quad
{\rm if} \quad c^{(1)} = 0 \ .
\ee

$\mu \sim \lambda_u\lambda_d M $ is the only non-trivial input parameter of the Higgs potential of the model
(which will be set $\sim m_{\rm weak}$ by the electroweak symmetry breaking).
Because we can take $\lambda_{u,d}\ll \alpha_{SM}$ the contributions
of the Higgs couplings to the soft SUSY-breaking terms are small
\be
B_\mu \ll \mu^2 \quad , \quad \delta m^2_{{u,d}} \sim 0 \sim \delta a^2_{u,d}
\label{diffsc2}
\ee
Therefore, these terms are essentially
the same as for
pure gauge mediation,
and the
prediction for scalar masses \cite{Meade:2008wd}
\be
\label{smasses}
m_{{u}}^2\, =\, m_{{d}}^2\, =\, m_{\tilde L}^2
\ee
is unchanged.

If we allow $M_{mess}$ and $M$ to be independent then we can easily
accommodate a non-vanishing $c^{(1)}$:
\be
\label{c1}
{B_\mu \over \mu^2} \, \sim \, {1 \over \lambda_u\lambda_d}\, {F \over M^2}
\sim \left(\frac{F}{\mu M_{mess}}\right)\left(\frac{M_{mess}}{M}\right)
\sim 16\pi^2 \left(\frac{M_{\tilde{\lambda}}}{\mu}\right)\left(\frac{M_{mess}}{M}\right)\,  \quad
{\rm if} \quad c^{(1)} \neq 0 \ .
\ee
This can clearly be chosen to be either of order one or much less than one.
Even in the former case one can then ensure that the scalar masses are really dominated by the true
GGM contributions by again choosing $\lambda_{u,d}\ll \alpha_{SM}$.
Most of the parameter space has negligibly small $B_{\mu}/\mu^2$ and values of order
one correspond to a tuning; our phenomenological analysis covers the generic
regions of parameter space where $B_{\mu}/\mu^2\ll \alpha^2_{SM}$.

\section{Spectra}
\label{sec:Spectra}

\begin{table}
\begin{center}
\begin{tabular}{|c|c|c|c|} \hline
  & $M_{mess}=10^6$ & $M_{mess}=10^{10}$ & $M_{mess}=10^{14}$ \\ \hline\hline
$\Lambda_G$~(GeV) & $1.08\times 10^5$ & $2.25\times 10^5$ & $1.86\times 10^5$ \\ \hline
$\Lambda_S$~(GeV) & $1.78\times 10^5$ & $4.96\times 10^3$ & $1.11\times 10^3$ \\ \hline\hline
 $\chi_1^0$ & {\bf 146} & 299 & 245               \\ \hline
$\chi_2^0$ & 286 & 547  & 462 \\ \hline
$\chi_3^0$ & 706 & 644 & 668 \\ \hline
$\chi_4^0$ & 712 & 677 & 683 \\ \hline
$\chi_1^{\pm}$ &  719 & 543 & 460\\ \hline
$\chi_2^{\pm}$ & 892 & 678 & 682 \\ \hline
$\tilde{g}$ & 1782 & 1534 & 1299       \\ \hline\hline

$\tilde{e}_L,\tilde{\mu}_L$ & 625 & 331 & 358 \\ \hline
$\tilde{e}_R,\tilde{\mu}_R$ & 312 & 135 & 172\\ \hline
$\tilde{\tau}_L$ & 633 & 344 & 365 \\ \hline
$\tilde{\tau}_R$ & 265 & {\bf 100} & {\bf 144} \\ \hline
$\tilde{\nu}_{1,2}$ & 620 & 324  & 352 \\ \hline
$\tilde{\nu}_3$ & 616 & 324 & 350 \\ \hline\hline

$\tilde{t}_1$ & 1649 & 1211 & 1108\\ \hline
$\tilde{t}_2$ & 1516 & 1059 & 924 \\ \hline
$\tilde{b}_1$ & 1650 & 1165 & 1060\\ \hline
$\tilde{b}_2$ & 1550 & 1204 & 1100\\ \hline
$\tilde{u}_1,\tilde{c}_1$ & 1822 & 1277 & 1182\\ \hline
$\tilde{u}_2,\tilde{c}_2$ & 1690 & 1215 & 1114\\ \hline
$\tilde{d}_1,\tilde{s}_1$ & 1784 & 1246 & 1154\\ \hline
$\tilde{d}_2,\tilde{s}_2$ & 1679 & 1213 & 1110\\ \hline\hline

$h_0$ & 117 & 117 & 117\\ \hline
$A_0, H_0$ & 569 & 722 & 740 \\ \hline
$H^{\pm}$ & 575 & 727 & 744 \\ \hline\hline
\end{tabular}
\caption{Best fit spectra for the three messenger scales. All masses are in GeV. The NLSP is shown in bold in each case.
(The LSP is the gravitino.)}
\end{center}
\label{fig:spectra}
\end{table}

Figure \ref{fig:spectra} shows the sparticle spectrum we obtain at the best-fit points of our scans. For $M_{mess}=10^6$~GeV the NLSP is the
neutralino with a mass of 146~GeV while the NNLSP is the lightest stau of mass 265~GeV. The gluino mass is approximately 1.8~TeV.
The lighter right-handed sleptons have masses around 300~GeV and the left-handed sleptons just over 600~GeV. As is usual for GMSB spectra, there is a
hierarchy between the sleptons and squarks. The average squark mass is 1670~GeV.

The spectrum for $M_{mess}=10^{10}$~GeV differs from the $M_{mess}=10^6$~GeV case in having lighter scalars  due the low value of $\Lambda_S$ and a more compressed gaugino spectrum.
The NLSP is the right-handed stau which just evades the bound of $98$~GeV from searches from CHAMPS at OPAL~\cite{Abbiendi:2003yd} and the NNLSP is the
degenerate selectron-smuon pair. The heavier sleptons still only have masses of just above 300~GeV.
The lightest gauginos are the neutralino and the chargino, and the gluino mass is 1.5~TeV. The squark masses are lighter in than the $M_{mess}=1\times 10^6$~GeV case by nearly 500~GeV and their average mass is 1.2~TeV.

There is a close resemblence between the $M_{mess}=1\times 10^{14}$~GeV and $M_{mess}=1\times 10^{10}$~GeV best-fit points. Both have a very large hierarchy $\Lambda_G \gg \Lambda_S$. For $M_{mess}=1\times 10^{14}$~GeV the NLSP is again the stau which in this case is slightly heavier than that for the best-fit point, and the NNLSP is the selectron-smuon pair. As the value of $\Lambda_G$ is slightly smaller, the squarks are correspondingly lighter.

\bibliographystyle{utphys}
\bibliography{durham}

\end{document}